\documentclass[useAMS,usenatbib]{mn2e}

\usepackage{epsfig}
\usepackage{aas_macros}
\usepackage{natbib}
\usepackage{times}

\newbox\grsign \setbox\grsign=\hbox{$>$} \newdimen\grdimen \grdimen=\ht\grsign
\newbox\labox \newbox\gabox \newbox\simpropbox \newbox\wtildebox 
\setbox\gabox=\hbox{\raise.5ex\hbox{$>$}\llap
    {\lower.5ex\hbox{$\sim$}}}\ht1=\grdimen\dp1=0pt
\setbox\labox=\hbox{\raise.5ex\hbox{$<$}\llap
    {\lower.5ex\hbox{$\sim$}}}\ht2=\grdimen\dp2=0pt
\def\ga{\mathrel{\copy\gabox}}
\def\la{\mathrel{\copy\labox}}

\newcommand{\msun}{\mbox{M$_\odot$}}
\newcommand{\dau}{\partial}
\newcommand{\e}{{\rm e}}
\newcommand{\F}{{\cal F}}
\newcommand{\be}{\begin{equation}}
\newcommand{\ee}{\end{equation}}
\newcommand{\bea}{\begin{eqnarray}}
\newcommand{\eea}{\end{eqnarray}}

\title{An uncertainty principle for star formation. I. Why galactic star formation relations break down below a certain spatial scale}
\author{J.~M.~Diederik Kruijssen$^1$ and Steven~N.~Longmore$^2$\\
$^{1}$Max-Planck Institut f\"{u}r Astrophysik, Karl-Schwarzschild-Stra\ss e 1, 85748 Garching, Germany; kruijssen@mpa-garching.mpg.de\\
$^{2}$Astrophysics Research Institute, Liverpool John Moores University, IC2, Liverpool Science Park, 146 Brownlow Hill,
Liverpool L3 5RF, United Kingdom}

\begin{document}

\date{Accepted 2014 January 15.  Received 2013 November 8; in original form 2013 July 9.}

\pagerange{\pageref{firstpage}--\pageref{lastpage}} \pubyear{2013}
\label{firstpage}

\maketitle

\begin{abstract}
Galactic {scaling relations between the (surface densities of) the gas mass and the star formation (SF) rate} are known to develop substantial scatter or {even change form} when considered below a certain spatial scale. We quantify how this behaviour should be expected due to the incomplete statistical sampling of independent star-forming regions. Other {included} limiting factors are the incomplete sampling of SF tracers from the stellar initial mass function and the spatial drift between gas and {stars}. We present a simple uncertainty principle for SF, which can be used to predict and interpret the failure of galactic SF relations on small spatial scales. {This uncertainty principle explains how the scatter of SF relations depends on the spatial scale and predicts a scale-dependent bias of the gas depletion time-scale when centering an aperture on gas or SF tracer peaks.} We show how the scatter and bias are sensitive to the physical size and time-scales involved in the SF process {(such as its duration {or the molecular cloud lifetime})}, and illustrate how our formalism provides a powerful tool to constrain these largely unknown quantities. {Thanks to its general form, the uncertainty principle can also be applied to other astrophysical systems, e.g.~addressing the time-evolution of star-forming cores, protoplanetary discs, or galaxies and their nuclei.}
\end{abstract}


\section{Introduction} \label{sec:intro}
Galactic star formation (SF) relations \citep[e.g.][]{silk97,kennicutt98b,elmegreen02,bigiel11} break down below a certain spatial scale \citep{schruba10,onodera10,liu11}. This failure of SF relations to describe small scales may potentially provide a better understanding of galactic-scale SF than the relations themselves do \citep[e.g.][]{blanc09,schruba10,calzetti12,leroy12,leroy13}.

SF relations should be expected to break down at some point -- they relate the gas and radiation properties of the SF process, which on some scale should no longer correlate because they cover subsequent phases of SF. For instance, a gas-free, young stellar cluster is detectable in SF tracers such as H$\alpha$ or UV emission, but not in the gas tracers that were likely visible at an earlier stage. This example can be expressed in terms of the statistical sampling of the SF process -- galactic SF relations average over all phases and implicitly assume that each phase is statistically well-sampled.

In this paper, we present a simple uncertainty principle that is similar in form to the famous criterion of \citet{heisenberg27}. It can be used to identify the spatial scale on which SF relations break down due to the incomplete sampling of the SF process, as well as to quantify the resulting scatter around and bias of such relations. After introducing our {framework}, we illustrate its use with a number of idealised examples. We show that it accurately describes the observed range of spatial scales on which the SF relations can be applied, as well as their scatter {and bias on} smaller scales. {It is also shown how the uncertainty principle can be used to derive a number of fundamental characteristics of the SF process.} {We validate the method using Monte-Carlo models of star-forming regions in galaxies.} In a {follow-up} paper \citep[hereafter K14]{kruijssen14}, we apply the uncertainty principle to observational data.

{At http://www.mpa-garching.mpg.de/KL14principle, we have made Fortran and IDL modules for applying the uncertainty principle publicly available. A checklist detailing the required steps for their observational application is supplied in Appendix~\ref{sec:appcode}.}

\section{An uncertainty principle for star formation}  \label{sec:principle}

\subsection{The general uncertainty principle} \label{sec:general}
We adopt the hypothesis that SF relations only apply above certain spatial scales because on smaller scales the different phases of the SF process are statistically not well-sampled. This may lead to the perception of episodicity or a deviation from these SF relations: given an ensemble of star-forming regions, certain permutations of their states will lead to SFRs higher or lower than predicted by SF relations. {For instance, if all regions in the ensemble happen to be on the verge of initiating SF, they will have very low SFRs for their gas masses in comparison to the galactic average.} The appropriate sampling of the phases of the SF process can be achieved by increasing the sample of states, or by somehow following the different phases in time. If a well-sampled SF relation is achieved by covering a large area or volume, this implies that the size-scale $\Delta x$ on which the SF relation is evaluated should satisfy:
\be
\label{eq:dx}
\Delta x \geq A\lambda ,
\ee
where $A$ and $\lambda$ represent a {dimensionless} constant and a length-scale, respectively, both of which are specified below.

The fact that a single star-forming region is observed at a specific time implies that that region by itself cannot satisfy galactic SF relations. However, if it were followed in time for the full duration of the SF process $\tau$ such that all `relevant' phases (i.e.~those that are traced in galactic SF relations) of the SF process are covered, {then the time-averaged properties of the region should be consistent with the galactic SF relations -- provided that the physics of SF are universal}. Specifically, to retrieve the SF relation for a single star-forming region, we require the time-scale over which the properties of the region are averaged {to cover at least} the time $\tau$ covering all phases of the SF process traced in galactic SF relations. Observationally, {this condition} can never be satisfied -- we cannot observe a stellar cluster and its progenitor cloud at the same time. {This is why galactic SF relations must} consider spatial scales large enough to cover multiple star-forming regions.

Observationally, {each phase of the SF process is probed with a different tracer (e.g.~CO for molecular gas, H$\alpha$ for SF). The emission from these tracers is observable for some fraction of the total SF process, although an individual observation only retrieves a snapshot at a discrete moment in time.} The detection of a gas or SF tracer does not distinguish at which time along the phase probed by that tracer it is observed. For instance, the gas mass that will eventually {participate in the SF process} is visible for a certain duration and can be detected throughout. {{If we now consider an ensemble of star-forming regions at randomly distributed times along the SF process}, the key consequence is that} a galactic SF relation can only be retrieved if the shortest phase of the SF process is sampled at least once. {In other words}, the phase with the shortest duration is the limiting factor in whether or not the galactic SF relation is retrieved. We define the duration of the SF process as
\be
\label{eq:tau}
\tau=\sum_{i=1}^N t_{{\rm ph},i}-\sum_{i=1}^{N-1} t_{{\rm over},i,i+1} ,
\ee
where $t_{{\rm ph},i}$ is the duration of phase $i$ and $t_{{\rm over},i,i+1}$ represents the duration of the overlap between phases $i$ and $i+1$. Defining $\Delta t\equiv\min{(t_{{\rm ph},i})}$ as the shortest phase of the SF process that is traced in galactic SF relations, the ratio $\tau/\Delta t$ then reflects the number of {\it independent} star-forming regions $N_{\rm indep,req}$ that need to be sampled in order to retrieve the statistically converged, galactic SF relation. We illustrate this result with an example. If the SF were to consist of two phases such that $t_{\rm ph,1}/t_{\rm ph,2}=9$ (i.e.~90~per~cent of the time is spent in the first, e.g.~molecular gas, phase), then $\Delta t/\tau=0.1$ and hence $N_{\rm indep,req}=10$ independent star-forming regions need to be covered to retrieve the galactic SF relation.

{The size of an `independent' region refers to {\it the largest spatial scale on which SF events within that region are correlated}, e.g.~by the global gravitational collapse of a molecular cloud, triggered SF, or a galaxy-scale perturbation such as a merger.} The number of independent star-forming regions with a characteristic {separation} $\lambda$ that is sampled within a two-dimensional aperture of diameter $\Delta x$ is $N_{\rm indep}=(\Delta x/\lambda)^2$. {The condition that this number exceeds the number of independent regions {\it required} to retrieve the galactic SF relation ($N_{\rm indep}\geq N_{\rm indep,req}$) thus yields}:
\be
\label{eq:nindep}
\left(\frac{\Delta x}{\lambda}\right)^{2} \geq \frac{\tau}{\Delta t} .
\ee
This can be rewritten in the familiar form of an uncertainty principle that needs to be satisfied for galactic SF relations to hold:
\be
\label{eq:dxdt}
\Delta x\Delta t^{1/2} \geq \lambda\tau^{1/2} .
\ee
Here, $\Delta x$ is the spatial scale {on which the SF relations are measured} and $\Delta t$ is the duration of the shortest phase of the SF process that is traced by the SF relation in question. {If equation~(\ref{eq:dxdt}) is satisfied, then the shortest phase of the SF process is {expected to be} well-sampled {(modulo Poisson noise)}, and hence the galactic SF relation is retrieved. Given the values of all $t_{{\rm ph},i}$, this defines the minimum size-scale:}
\be
\label{eq:dxsamp}
\Delta x \geq \Delta x_{\rm samp}\equiv\left(\frac{\tau}{\Delta t}\right)^{1/2}\lambda ,
\ee
{which specifies the constant $A$ in equation~(\ref{eq:dx}).} {Hence, (1) the more similar the various $t_{{\rm ph},i}$ are or (2) the smaller the size is of independent regions, the smaller the minimum size-scale is on which galactic SF relations still hold.}

Next to the statistical sampling in time and space of the full SF process, additional scatter is introduced by the incomplete sampling of the SF tracer at low SFRs \citep{lee11,fumagalli11}. Observational estimates of the SFR almost exclusively rely on emission from massive stars, which statistically may not be produced at low SFRs, {leading} to an underestimation of the SFR. If we define a minimum ${\rm SFR}_{\rm min}$ {above which a certain SF tracer arises from an adequately sampled} IMF (see Table~\ref{tab:tracers}), a given SFR surface density $\Sigma_{\rm SFR}$ implies that a spatial scale of
\be
\label{eq:dximf}
\Delta x\geq \Delta x_{\rm IMF}\equiv\left(\frac{4}{\pi}\frac{{\rm SFR}_{\rm min}}{\Sigma_{\rm SFR}}\right)^{1/2}
\ee
is required to retrieve a reliable SFR estimate. Here, the factor of four enters because $\Delta x$ represents a diameter rather than a radius.

Finally, the above limits on the spatial scale on which galactic SF relations hold only apply if the {relative} flux of gas and SF tracers across the boundary of an aperture is negligible over the duration of the SF process $\tau$. Any net drift will introduce scatter or potentially systematic deviations. Given a {characteristic drift velocity} $\sigma$, the size-scale should therefore satisfy
\be
\label{eq:dxdrift}
\Delta x\geq \Delta x_{\rm drift}\equiv\frac{1}{2}\sigma \tau ,
\ee
where the factor of $1/2$ arises from taking the {expected time difference between two regions positioned at random times during} the gas and stellar phases. In practice, the condition of equation~(\ref{eq:dxsamp}) implies that multiple independent star-forming regions are covered in a single aperture. Statistically speaking, the flux of gas and SF tracers across the aperture boundary is therefore {much smaller than across individual star-forming regions} and equation~(\ref{eq:dxdrift}) should typically be satisfied. This will be illustrated in \S\ref{sec:examples} below.

Together, the above three conditions postulate that galactic SF relations hold on size-scales
\be
\label{eq:dxtot}
\Delta x \geq \Delta x_{\rm max}\equiv\max\{\Delta x_{\rm samp},\Delta x_{\rm IMF},\Delta x_{\rm drift}\} .
\ee

\subsection{Specification of key variables} \label{sec:specific}
The general uncertainty principle of equation~(\ref{eq:dxtot}) should be satisfied when evaluating galactic SF relations, irrespective of the precise choice of the variables it depends on. However, the practical application of the principle requires the specification of the (tracer-dependent) durations of the several phases of the SF process $t_{{\rm ph},i}$, the minimum SFRs required for their use ${\rm SFR}_{\rm min}$, the total duration of the SF process $\tau$, and the characteristic spatial separation of independent star-forming regions $\lambda$. {Note that the numbers in this section are strictly adopted for the purpose of illustration. In practice, they can be measured from the observational data (see \S\ref{sec:examples} and \S\ref{sec:future}).}

\begin{figure}
\center\resizebox{\hsize}{!}{\includegraphics{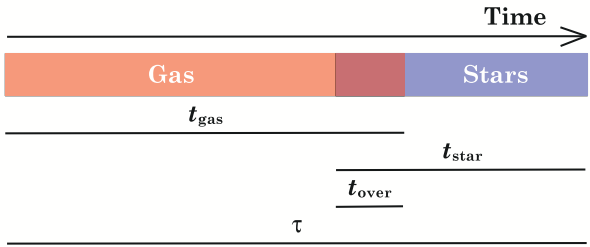}}\\
\caption[]{\label{fig:tschem}
      {Schematic representation of $t_{\rm gas}$, $t_{\rm star}$, $t_{\rm over}$ and $\tau$. {Depending on the adopted gas and SF tracers {(and their detectability)}, $t_{\rm over}$ can refer to the duration of SF itself, the duration of gas removal, or a combination of both (see text).}}
                 }
\end{figure}
In the rest of this paper, we illustrate the use of our uncertainty principle by {assuming that the SF process as traced by galactic scaling relations consists of two steps, as is shown schematically in Figure~\ref{fig:tschem}}. During the first phase, the gas tracer can be detected while no SF is seen, whereas during the second phase, a young stellar population is in place and the gas has been expelled due to stellar feedback (or any other mechanism that leads the gas phase to become undetectable in the adopted gas tracer). The duration of the entire SF process is therefore given by $\tau=t_{\rm gas}+t_{\rm star}-t_{\rm over}$. Here, $t_{\rm gas}$ denotes the duration of the first phase, {i.e.~the time for which the gas is visible in the adopted tracer before entering the SF process, including possible interruptions}, $t_{\rm star}$ represents the duration of the second phase, and $t_{\rm over}$ is the duration of the overlap between the gas and stellar phases. As a result, $\Delta t=\min\{t_{\rm gas},t_{\rm star}\}$, and in the first examples we shall set $t_{\rm over}=0$. In reality, more phases {may} exist and {inevitably} $t_{\rm over}\neq 0$ (see \S\ref{sec:physics}).

{The meaning of $t_{\rm over}$ depends on the adopted gas and SF tracers. If a {diagnostic} is used that traces SF only in the unembedded state (such as H$\alpha$), then $t_{\rm over}$  {may be dominated by} the time it takes the stellar feedback to remove the gas from the aperture {(either by a phase transition or by motion)}. If a {diagnostic} is used that traces SF from the onset of SF, even when it is still deeply embedded (such as cm continuum or young stellar object counts), then $t_{\rm over}$ reflects the duration of SF itself plus the time-scale for gas removal. Inevitably, the precise onset of the overlap then depends on the sensitivity to the SF tracer and the time-evolution of the star formation efficiency {(or feedback)}. For several of the examples below, we assume that during the overlap the gas and SF tracers are {\it statistically} detected at 50~per~cent of their normal intensity, consistent with a linear decrease (gas) or increase (stars). Higher-order functional forms \citep[e.g.][]{burkert13} {are likely more accurate, but these typically introduce corrections of only a few 10~per~cent, which justifies the use of a linear form}.} {However, note that the equations provided below are generalised and do not depend on the detailed intensity evolution of the gas and SF tracers.} The overlap can only be ignored if (1) its duration is much shorter than that of the individual gas and SF tracers and (2) multiple independent star-forming regions are covered {in the aperture}. We illustrate the effect of a non-zero overlap in \S\ref{sec:examples}.

\begin{table}
 \centering
  \begin{minipage}{70mm}
  \caption{Example star formation tracer properties}\label{tab:tracers}
  \begin{tabular}{@{}l c c c c c@{}}
  \hline 
 Tracer             & ${\Delta t_{50\%}}^a$ & ${\Delta t_{95\%}}^a$ &  ${\langle t\rangle_{\rm lum}}^a$ & SFR$_{\rm min}$ \\
                          & (Myr)                       & (Myr)                         & (Myr)                                          & (10$^{-3}~\msun~{\rm yr}^{-1}$) \\ \hline
 H$\alpha$  &    1.7                        &  4.7                           &  2                                                & 1$^b$\\
 FUV                 &    4.8                        &  65                            & 14                                               &  0.04$^c$\\
  \hline
\end{tabular}\\
$^a$From \citet{leroy12}.
$^b$From \citet{kennicutt12}.
$^c$Assuming a FUV flux contribution from stars $M\geq3~\msun$.
\end{minipage}
\end{table}
For {two} popular SF tracers, Table~\ref{tab:tracers} shows times at which 50~and 95~per~cent of the total flux has been emitted for a typical stellar population \citep[see][]{leroy12}, as well as the luminosity-weighted durations $\langle t\rangle_{\rm lum}$ of each tracer and the minimum SFR required for adequately sampling the traced stars from a standard \citet{chabrier03} IMF. For the applications below, we set $t_{\rm star}$ equal to the luminosity-weighted duration, {although in practice other choices may be more appropriate depending on the observational sensitivity limits}.
\begin{table*}
 \centering
  \begin{minipage}{164mm}
  \caption{Adopted properties of idealised galaxies and regions.}\label{tab:examples}
  \begin{tabular}{@{}l c c c c c c c c c c@{}}
  \hline 
  Galaxy/ & $\Sigma$ & $\Omega_{-2}$ & $\sigma$ & $\Sigma_{\rm SFR}$ & $t_{\rm gas}$ & $l_{\rm T}$ & $\Delta x_{\rm samp}$ & $\Delta x_{\rm IMF}$ & $\Delta x_{\rm drift}$ & $\Delta x_{\rm max}$ \\
  region & ($\msun$~pc$^{-2}$) & (100~Myr)$^{-1}$ & (km~s$^{-1}$) & ($\msun$~yr$^{-1}$~kpc$^{-2}$) & (Myr) & (kpc) & (kpc) & (kpc) & (kpc) & (kpc) \\
 \hline
SN & 10 & 2.6 & 7 & 0.0063 & 38 & 0.21 & {0.94} & 0.45 & 0.14 & 0.94 \\
CMZ & 120 & 85 & 35 & 0.20 & 1.2 & 0.0023 & 0.0039 & {0.080} & 0.057 & 0.08 \\
Disc & 15 & 4.0 & 10 & 0.011 & 25 & 0.13 & {0.49} & 0.34 & 0.14 & 0.49 \\
Dwarf & 10 & 3.0 & 10 & 0.0063 & 33 & 0.16 & {0.66} & 0.45 & 0.18 & 0.66 \\
SMG & $10^3$ & 30 & 50 & 20 & 3.3 & 0.16 & {0.26} & 0.0080 & 0.14 & 0.26 \\
\hline
\end{tabular}\\
The solar neighbourhood (SN) values are based on a flat $V_{\rm c}=220~{\rm km}~{\rm s}^{-1}$ rotation curve and the \citet{wolfire03} gas model; the Central Molecular Zone (CMZ) values are taken from \citet[230~pc-{\it averaged}]{kruijssen13}; the disc and dwarf galaxy values are based on \citet{leroy08}; the sub-mm galaxy (SMG) values are based on \citet{genzel10}. Except for the CMZ and the SMG, the SFR surface densities assume a \citet{kennicutt98b} SF relation. Boldface $\Delta x$ values indicate the maxima of equation~(\ref{eq:dxtot}). In all cases, $t_{\rm star}=2$~Myr is adopted (see text).
\end{minipage}
\end{table*}

The duration of the gas phase is not well-constrained, and depends on the specific gas tracer as well as the galactic environment. If a molecular gas tracer is used in a Milky~Way-like galaxy, {a plausible} duration of its detectability is the dynamical time-scale of the host galaxy, which sets the time interval between external perturbations such as cloud-cloud collisions or spiral arm passages. By contrast, if a tracer of atomic gas is used, then the condensation time-scale to the molecular form enters. In the following examples, we assume the use of a molecular gas tracer, and set $t_{\rm gas}=\Omega^{-1}$, where $\Omega\equiv V/R$ is the angular velocity at circular velocity $V$ and galactocentric radius $R$. {In galaxy discs in hydrostatic equilibrium, this time-scale is similar to the free-fall time of GMCs \citep[e.g.][]{krumholz05}. However, equilibrium is not always satisfied, and therefore} an alternative definition would be to use the typical {observed} free-fall time of GMCs. As shown in \S\ref{sec:examples} and \S\ref{sec:future}, our uncertainty principle can actually be used to empirically determine the time-scales during which gas tracers are detectable.

In galactic discs, a good proxy for the typical separation of independent star-forming regions $\lambda$ is the Toomre length, which for a flat rotation curve is given by
\be
\label{eq:ltoomre}
l_{\rm T}=\frac{\pi G\Sigma}{\Omega^2} ,
\ee
where $\Sigma$ is the gas surface density. Substituting the above expressions for $\tau$ and $\lambda$ in equation~(\ref{eq:dxsamp}), we can specify the spatial scales above which galactic SF relations apply due to the statistical sampling of independent star-forming regions
\be
\label{eq:dxsamp2}
\Delta x_{\rm samp} = \left(\frac{\Omega^{-1}+t_{\rm star}-t_{\rm over}}{\min\{\Omega^{-1},t_{\rm star}\}}\right)^{1/2}\frac{\pi G\Sigma}{\Omega^2} ,
\ee
and due to drift
\be
\label{eq:dxdrift2}
\Delta x_{\rm drift} = \frac{1}{2}\sigma(\Omega^{-1}+t_{\rm star}-t_{\rm over}),
\ee
where $\sigma$ is assumed to be the gas velocity dispersion measured on the largest scale {young stars are assumed to inherit the velocity dispersion of their natal cloud)}. Equation~(\ref{eq:dximf}) does not require to be specified further. The above equations show that the gas surface density $\Sigma$, the SFR surface density $\Sigma_{\rm SFR}$, the rotation curve, and the gas velocity dispersion $\sigma$ must be known to apply our uncertainty principle in the way specified here. Galactic environments other than discs (such as galaxy mergers) may require different definitions for $t_{\rm gas}$, $\tau$ and $\lambda$.

\section{Application to idealised examples} \label{sec:examples}
\subsection{The failure of SF relations on small spatial scales}
We now illustrate the use of the uncertainty principle quantitatively with a number of archetypical galactic environments, which are listed in Table~\ref{tab:examples} together with their characteristic properties and the resulting $\Delta x_i$. We consider the solar neighbourhood (SN), the Central Molecular Zone (CMZ) of the Milky Way, a `disc galaxy', a `dwarf galaxy', and a starbursting `sub-mm galaxy' (SMG). Throughout this section, we will assume the use of H$\alpha$ to trace SF. Adopting the luminosity-weighted duration, we define $t_{\rm star}=2~{\rm Myr}$ (see Table~\ref{tab:tracers}).

In all cases other than the CMZ, the galactic SF relation breaks down first due to the incomplete sampling of independent star-forming regions (i.e.~below $\Delta x_{\rm samp}$), whereas in the CMZ the SFR is too low to adequately sample the SF tracer from the IMF on scales $l_{\rm ap}<80$~pc. {This is due to the low SFR measured in the CMZ \citep{longmore13}, but even if the CMZ were forming stars at a higher rate, the effect of drift would still be more important than the incomplete sampling of star-forming regions (i.e.~$\Delta x_{\rm drift}>\Delta x_{\rm samp}$).} The typical size-scales above which galactic SF relations hold vary from $\sim100$~pc in the CMZ to almost a kpc in the SN. {It can be inferred from equation~(\ref{eq:dxsamp2}) that this latter value is representative for much of the population of nearby, star-forming galaxies. These galaxies satisfy a broad trend of $\Omega\propto\Sigma^{0.5}$ with about 0.3--0.5~dex scatter \citep{krumholz05,kruijssen12d} because they simultaneously satisfy the Schmidt-Kennicutt and Silk-Elmegreen SF relations \citep{schmidt59,silk97,elmegreen97b,kennicutt98b}. Substitution of this trend into equation~(\ref{eq:dxsamp2}) results in a roughly constant $\Delta x\sim1$~kpc for all surface densities, with a variation of 0.6--1.0 dex as in Table~\ref{tab:examples}.} Because the size-scales of GMCs and stellar clusters are much smaller \citep{longmore14}, it is clear that they cannot satisfy galactic SF relations.

\begin{figure}
\center\resizebox{\hsize}{!}{\includegraphics{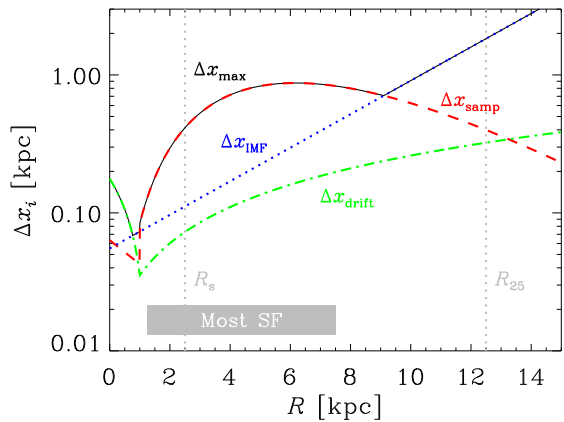}}\\
\caption[]{\label{fig:dx_rgal}
      {Example of the minimum size-scales $\Delta x_i$ above which galactic SF relations hold as a function of galactocentric radius in an idealised disc galaxy (see text). The {red} dashed line shows $\Delta x_{\rm samp}$, which accounts for the statistical sampling of the different phases of the SF process. The {blue} dotted line shows $\Delta x_{\rm IMF}$, which accounts for the sampling of the high-mass (SF-tracing) end of the IMF. The {green} dash-dotted line shows $\Delta x_{\rm drift}$, which accounts for the drift of gas and young stars across the aperture boundary. At each radius, the {black} solid line shows the maximum of these three limits, $\Delta x_{\rm max}$. {The vertical dotted lines indicate the scale radius $R_{\rm s}$ and the optical radius $R_{25}\sim5R_{\rm s}$ \citep[e.g.][]{leroy08}, while the grey box indicates the radius interval where most of the SF occurs in an exponential disc ($R=0.5$--$3~R_{\rm s}$).}}
                 }
\end{figure}
In Figure~\ref{fig:dx_rgal}, we illustrate the radial profiles of the three different $\Delta x_i$ from equations~(\ref{eq:dxsamp})--(\ref{eq:dxdrift}) as well as $\Delta x_{\rm max}$ for a disc galaxy with a logarithmic potential ($V=200~{\rm km}~{\rm s}^{-1}$), an exponential gas surface density profile with central value $\Sigma(0)=200~\msun~{\rm pc}^{-2}$ and scale radius $R_{\rm s}=2.5~{\rm kpc}$, and a gas velocity dispersion of $\sigma=10~{\rm km}~{\rm s}^{-1}$. In the central kpc, the velocity dispersion is assumed to rise linearly to a central value of $\sigma=50~{\rm km}~{\rm s}^{-1}$, and the rotation curve is assumed to be solid-body (i.e. $\Omega$ is constant). {The \citet{kennicutt98b} relation is used to translate the gas surface density to a SFR surface density.}

In the central 500~pc, the drift of gas and young stars across the aperture boundary set the minimum scale on which galactic SF relations hold ($\Delta x_{\rm drift}\sim150$~pc), which is consistent with the known offset between the dense gas and 24$\mu$m sources in the CMZ of the Milky Way \citep{yusefzadeh09,longmore13}. However, the incomplete sampling of independent star-forming regions sets $\Delta x_{\rm samp}=100$--$900$~pc over the largest part of the galaxy, in the radius range $R=1$--$9~{\rm kpc}$. At larger radii, the incomplete sampling of SF tracers from the IMF kicks in and $\Delta x_{\rm IMF}>1$~kpc. A direct implication is that when using SF tracers that {are visible over a longer age range} (e.g.~FUV), the IMF remains properly sampled out to larger radii. This is why discs do not show the same truncation in the UV \citep{thilker07} as when observed in H$\alpha$ (\citealt{martin01}; also see \citealt{bigiel10}). Considering that the majority of the SF in exponential disc galaxies occurs between 0.5 and 3 gas scale radii, Figure~\ref{fig:dx_rgal} shows that $\Delta x_{\rm samp}$ most strongly restricts the application of galactic SF relations on small scales.

\subsection{The scatter of SF relations} \label{sec:scatter}
Galactic SF relations are often characterised by defining the gas depletion time-scale $t_{\rm depl}\equiv M_{\rm gas}/{\rm SFR}$, which allows the scatter of the SF relation to be quantified as the root-mean-square (RMS) scatter of the depletion time-scale $\sigma_{\log{t}}$ \citep[e.g.][]{bigiel08,schruba11,leroy13}. In the framework of the uncertainty principle presented in \S\ref{sec:general}, the scatter of the SF relation for an aperture with diameter $l_{\rm ap}$ is determined by the Poisson statistics of the number of times the SF phases are sampled within the aperture, which will be dominated by the shortest phase. {Additional sources of scatter are the luminosity evolution of the gas and stars during $t_{\rm gas}$ and $t_{\rm star}$, respectively, the mass spectrum of the independent regions, and the intrinsic observational error:
\be
\label{eq:scatter}
\sigma_{\log{t}}^2=\sigma_{\rm samp}^2+\sigma_{\rm evo}^2+\sigma_{\rm MF}^2+\sigma_{\rm obs}^2 ,
\ee
where $\sigma_{\rm samp}$ indicates the Poisson error, $\sigma_{\rm evo}$ represents the scatter caused by the luminosity evolution of independent regions, $\sigma_{\rm MF}$ is the scatter due to the mass spectrum, and $\sigma_{\rm obs}$ denotes the intrinsic observational error. The full derivation of these four components is presented in Appendix~\ref{sec:appscatter} and we describe their qualitative behaviour here.

For $l_{\rm ap}\gg\lambda$, the scatter due to the Poisson statistics of sampling independent regions decreases with aperture size as $\sigma_{\rm samp}\propto l_{\rm ap}^{-1}$, because the relative Poisson error of the number of regions covered in an aperture is $\sigma_{\ln{N}}=\sigma_N/N=N^{-1/2}=\lambda/l_{\rm ap}$. The distribution of these $N$ regions over gas and stellar regions is set by the fractions $t_{\rm gas}/\tau$ and $t_{\rm star}/\tau$, respectively, implying that for large $l_{\rm ap}$ we have:}
\be
\label{eq:scatter2}
\sigma_{\log{t}}=\alpha\left[1+\min{\left(\frac{t_{\rm star}}{t_{\rm gas}},\frac{t_{\rm gas}}{t_{\rm star}}\right)}\right]^{1/2}\frac{\Delta x_{\rm max}}{l_{\rm ap}} ,
\ee
where $\alpha\equiv1/\ln{10}\approx0.43$ {converts the logarithmic scatter from base $\e$ to base $10$}, the term in brackets accounts for the conversion of the scatter of a single tracer to the combined scatter of both tracers on $t_{\rm depl}$, and the ratio $\Delta x/l_{\rm ap}$ counts the number of shortest SF phases within the aperture.\footnote{We have tested the influence of a non-homogeneous environment (e.g.~an exponential disc galaxy with a scale radius $R_{\rm s}\la l_{\rm ap}$) and the resulting variation of $\Delta x_{\rm max}$ within the aperture. We find that equation~(\ref{eq:scatter2}) remains accurate to within a few per cent under all physical circumstances.} {If the apertures were truly randomly positioned, this expression would describe the scatter for all spatial scales. However, the scatter does not keep increasing indefinitely towards small aperture sizes ($l_{\rm ap}<\lambda$), because only apertures that contain both the gas and SF tracer are included --  otherwise the gas depletion time-scale would be zero or infinity. If $t_{\rm over}=0$, this selection bias implies that very small apertures each typically contain the bare minimum of one gaseous region and one stellar region (the probability of catching more is negligible), which causes the scatter to vanish. This means that at small aperture sizes, the Poisson scatter actually increases with aperture size rather than the decrease that is seen at $l_{\rm ap}\gg\lambda$. In this part of the size-scale range, the luminosity evolution and the mass spectrum therefore dominate the scatter. In the simple case of $t_{\rm gas}=t_{\rm star}$, the transition between both regimes occurs at $l_{\rm ap}=2$--$3\lambda$.

The other terms of equation~(\ref{eq:scatter}) will vary substantially between different tracers, galaxies, and specific observations. The scatter due to the luminosity evolution of gaseous regions during $t_{\rm gas}$ and stellar regions during $t_{\rm star}$ depends on its particular functional form. Likewise, the scatter due to the mass spectrum of the independent regions also depends on its detailed characteristics. In both cases, the scatter does decrease with the aperture size, simply because the scatter of the mean decreases with the number of regions sampled, i.e.~$\sigma_{\rm mean}=\sigma_1 N^{-1/2}$, where $\sigma_1$ is the scatter for a single region. As stated previously, the number of sampled regions $N\propto(l_{\rm ap}/\lambda)^2$ for $l_{\rm ap}\gg\lambda$, but for small apertures the selection bias of requiring the presence of both tracers prevents $N<1$ (if $t_{\rm over}\neq0$) or $N<2$ (if $t_{\rm over}=0$). Detailed expressions are again provided in Appendix~\ref{sec:appscatter}, and are included in the publicly available routines (see Appendix~\ref{sec:appcode}). We leave the scatter due to the luminosity evolution and the mass spectrum for single regions as a free parameter. Reasonable values are $\sigma_{\rm evo,1g}=\sigma_{\rm evo,1s}=0.3$~dex for the luminosity evolution of a single gaseous or stellar peak, respectively \citep[cf.][]{leroy12}, and $\sigma_{{\rm MF},1}\sim0.8$~dex, which is roughly appropriate for a power law mass spectrum with a slope of $-2$ over a factor of 40 in mass. Finally, the observational scatter acts as a constant lower limit over all size-scales, which in practice flattens the $\sigma_{\log{t}}$--$l_{\rm ap}$ relation at large aperture sizes.
}

\begin{figure}
\center\resizebox{\hsize}{!}{\includegraphics{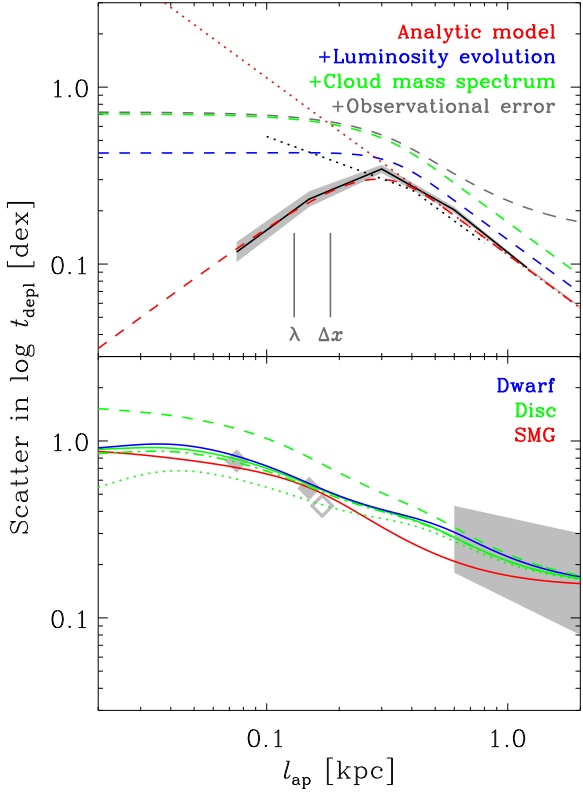}}\\
\caption[]{\label{fig:scatter}
      {Scatter in the gas depletion time-scale as a function of aperture size $l_{\rm ap}$. {\it Top panel}: The black solid line reflects the result of a simple Monte-Carlo experiment using 50,000 randomly placed apertures. The grey area indicates the RMS scatter of the scatter when using only 100 apertures. The red dashed line represents the analytic model from equation~(\ref{eq:scatter}) and Appendix~\ref{sec:appscatter} for $\sigma_{\rm evo}=\sigma_{\rm MF}=\sigma_{\rm obs}=0$, and the red dotted line indicates its power-law behaviour for $l_{\rm ap}\gg\lambda$ as expressed in equation~(\ref{eq:scatter2}). The \{blue,green,grey\} dashed lines subsequently add scatters of $\{\sigma_{\rm evo,1},\sigma_{\rm MF,1},\sigma_{\rm obs}\}=\{0.42,0.8,0.15\}$~dex to illustrate the effects of the luminosity evolution of independent regions, a cloud mass spectrum, and an intrinsic observational error, respectively. The black dotted line indicates the result from \citet{feldmann11} for a particular numerical setup (see text). {\it Bottom panel}:  The \{blue,green,red\} solid lines show the \{dwarf,disc,sub-mm\} galaxies from Table~\ref{tab:examples}. We have set $\sigma_{\rm evo,1g}=\sigma_{\rm evo,1s}=0.3$~dex, $\sigma_{\rm MF,1}=0.8$~dex and $\sigma_{\rm obs}=0.15$~dex as in the top panel. The \{dotted,dashed,dash-dotted\} lines refer to the disc galaxy model with $\sigma_{\rm MF,1}=0$~dex, $\sigma_{\rm MF,1}=1.6$~dex, and $\sigma_{\rm evo,1g}=\sigma_{\rm evo,1s}=0$~dex, respectively. The grey-shaded area indicates the part of parameter space covered by nine galaxies from the {\sc Heracles} survey \citep{leroy13}. The filled diamonds represent M33 \citep{schruba10} and the open diamond indicates M51 \citep{blanc09}.}
                 }
\end{figure}
{In addition to the analytic expression provided above, we estimate the scatter in $\log{t_{\rm depl}}$ with a simple Monte-Carlo experiment, in which we randomly distribute 50,000 points over an area such that their mean separation is $\lambda=130$~pc and position each region randomly on the time sequence of Figure~\ref{fig:tschem}, using time-scales $t_{\rm gas}=t_{\rm star}$ and $t_{\rm over}=0$. We do not include any possible evolution of each region's luminosity in either tracer, nor a region mass function or an intrinsic observational error. We then randomly place 50,000 apertures to measure the scatter in $\log{t_{\rm depl}}$, only including those apertures that include non-zero flux for both the gas and SF tracers. The resulting relation between the scatter and the aperture size is shown in the top panel of Figure~\ref{fig:scatter} (black line), which also illustrates that the variation of the obtained relation when using only 100 apertures is minor (grey area). At large aperture sizes, the scatter decreases as $\sigma_{\log{t}}\propto N^{-1/2}\propto l_{\rm ap}^{-1}$, which is expected for Poisson statistics. As explained above, the scatter also does not increase indefinitely towards small apertures, because apertures not including both tracers are discarded. For the adopted parameters (in particular $t_{\rm over}=0$), this requires at least two independent regions to be present in the aperture. The probability of finding more than that decreases rapidly when $l_{\rm ap}\ll\lambda$, implying that most apertures that are not discarded have the same content of one gaseous and one stellar region. As a result, the scatter goes to zero for $l_{\rm ap}\downarrow0$.

The top panel of Figure~\ref{fig:scatter} also shows that the analytic expression of equation~\ref{eq:scatter} and Appendix~\ref{sec:appscatter} (red dashed line) agrees very well with the Monte-Carlo experiment.\footnote{The small discrepancy at $l_{\rm ap}\sim2\lambda$ is not due to statistical noise and arises because the derivation in Appendix~\ref{sec:appscatter} does not include the covariance between the number of gaseous and stellar regions. The complete expression is considerably more complex, which is undesirable in view of the satisfactory accuracy of the presented form.} The decrease of the scatter at small aperture sizes vanishes when including $\sigma_{\rm evo,1g}=\sigma_{\rm evo,1s}=0.3$~dex scatter (blue dashed line). Even when all apertures have the same content of (at least) one stellar and one gaseous region, the evolution of the gas-to-stellar flux ratio leads to residual variance and hence the scatter approaches $\sigma_{\log{t}}^2=\sigma_{\rm evo,1g}^2+\sigma_{\rm evo,1s}^2$ for $l_{\rm ap}\downarrow0$. Similarly, including a scatter of $\sigma_{\rm MF}=0.8$~dex due to an underlying mass function (green dashed line) causes the scatter at small aperture sizes to saturate at $\sigma_{\log{t}}=\sqrt{\sigma_{\rm evo,1g}^2+\sigma_{\rm evo,1s}^2+\sigma_{\rm MF,1}^2/2}$. It also increases the scatter at large aperture sizes. Note that the presence or absence of a mass function does not affect the scatter at small aperture sizes when $t_{\rm over}\neq0$ -- the scatter at small aperture sizes is then dominated by single regions residing in the overlap phase\footnote{For typical parameters, this is more likely than catching one gaseous and one stellar region independently in a single aperture.}, for which a mass function affects the gas and stellar flux in the same way and hence does not introduce additional scatter. Finally, including an intrinsic observational error margin of $\sigma_{\rm obs}=0.15$~dex (grey dashed line) causes the scatter to saturate at $\sigma_{\log{t}}=\sigma_{\rm obs}$ for $l_{\rm ap}\rightarrow\infty$.

For reference, Figure~\ref{fig:scatter} includes the relation between the scatter and the aperture size that was found by \citet{feldmann11} in grid-based hydrodynamical simulations of (disc) galaxies. Although a detailed comparison is obstructed by the somewhat arbitrary position on the x-axis of each model (our models assume $\lambda=130$~pc), both results agree for $l_{\rm ap}\geq300$~pc. At smaller aperture sizes, the comparison is not representative because in that regime the size-scale dependence of the scatter depends on the details of the underlying luminosity evolution and the adopted mass spectrum. Nevertheless, the prediction that some flattening of the relation must occur below size-scales of a few 100~pc seems to be robust.}

Using equation~(\ref{eq:scatter}), we can now predict the scatter of the observed gas depletion time-scale as a function of aperture size for several of our example systems {from Table~\ref{tab:examples}}. This is shown in {the bottom panel of} Figure~\ref{fig:scatter} for the disc galaxy, dwarf galaxy and SMG parameter sets. {The galaxies follow roughly the same trend of decreasing scatter with aperture size, but there are several relevant differences. For instance, the bumps and slightly wave-like behaviour is caused by the dissimilar values of $t_{\rm gas}$ and $t_{\rm star}$ and the resulting increase of $\Delta x$ -- the bump visible at $l_{\rm ap}\sim500$~pc for the dwarf and disc galaxies coincides with $\Delta x=\sqrt{\tau/\Delta t}\lambda\sim 3.9\lambda\sim500$~pc. For $t_{\rm gas}\sim t_{\rm star}$, this bump would have moved to $l_{\rm ap}\sim\lambda$, as is the case for the SMG parameter set. The dotted, dashed and dash-dotted lines show how the size-scale dependence of the scatter depends on the luminosity evolution of individual regions and their underlying mass spectrum. While the variation is non-negligible, it clearly represents a secondary effect.} The overall trend is that the scatter varies from $\sigma_{\log{t}}\sim0.9$ at $l_{\rm ap}=50$~pc to $\sigma_{\log{t}}\sim0.2$ at $l_{\rm ap}=1$~kpc, {indicating a rough power-law relation of:
\be
\label{eq:scatter3}
\sigma_{\log{t}}\sim 0.2\left(\frac{l_{\rm ap}}{{\rm kpc}}\right)^{-0.5} ,
\ee
for $l_{\rm ap}=0.05$--$1$~kpc. Note that the details of this relation are by no means universal and should vary substantially between galaxies due to variations in $\lambda$, $t_{\rm gas}$, $t_{\rm over}$, $\sigma_{\rm evo}$ and $\sigma_{\rm MF}$. While a slope of $-1$ is expected for pure Poisson statistics, we see that a shallower slope can emerge due to the combined effect of the flattening at small aperture sizes that was explained above, the intrinsic observational error, and the dissimilarity of $t_{\rm gas}$ and $t_{\rm star}$.} Although a detailed comparison with observations is deferred to K14, {Figure~\ref{fig:scatter} does show which part of parameter space is covered by the observed nearby galaxies of \citet{leroy13}, as well as M33 \citep{schruba10} and M51 \citep{blanc09}, indicating that our model agrees very well with the trend of these observations.}

\subsection{How the uncertainty principle constrains SF physics} \label{sec:physics}
Thus far, we have assumed that apertures are randomly positioned on a galaxy. However, it is also possible to estimate the relative change of the measured gas depletion time-scale as a function of aperture size when centering it on a concentration of gas (increasing $t_{\rm depl}$ with respect to the galactic average) or young stars \citep[decreasing $t_{\rm depl}$ with respect to the galactic average, see][]{schruba10}. {In particular, we show in this section that this relative change (or bias) is a very useful quantity to constrain the time-scales governing the evolution of gas and SF in galaxies.}

{The relative change of the gas depletion time-scale when centering apertures on gas or stellar peaks can be estimated with a simple statistical model. The depletion time is defined as $t_{\rm depl}\equiv M_{\rm gas}/{\rm SFR}\propto \F_{\rm gas}/\F_{\rm SF}$, where $\F_{\rm gas}$ and $\F_{\rm SF}$ indicate the flux emitted by gas and SF tracers, respectively. The expected flux from both tracers in apertures focussed on gas or stellar peaks follows from the Poisson statistics of independent regions in apertures of varying size. The resulting flux ratio can then be compared to the galaxy-wide flux ratio to obtain the relative change of the measured gas depletion time-scale. The derivation is presented in detail in Appendix~\ref{sec:appbias} and only the result is provided here.}

{By centering an aperture on a gas peak, the gas flux is guaranteed to be non-zero and may increase due to additional gas-rich regions residing in the aperture by chance -- the expected number of these `contaminants' increases with the aperture size. By contrast, the stellar flux could potentially be zero, because it is constituted by the sum of the flux emitted by stellar regions residing in the aperture {\it by chance} and the flux emitted by the central gas peak if it happens to be in the overlap phase (which can only occur if $t_{\rm over}\neq 0$). The relative change of the gas depletion time-scale then becomes}
\be
\label{eq:tdeplgas}
\frac{\left[t_{\rm depl}\right]_{\rm gas}}{\left[t_{\rm depl}\right]_{\rm gal}}=
\frac{1+\frac{t_{\rm gas}}{\tau}\left(\frac{l_{\rm ap}}{\lambda}\right)^2}
       {\beta_{\rm s}\frac{t_{\rm over}}{t_{\rm star}}\left[1+(\beta_{\rm s}-1)\frac{t_{\rm over}}{t_{\rm star}}\right]^{-1}+\frac{t_{\rm gas}}{\tau}\left(\frac{l_{\rm ap}}{\lambda}\right)^2} ,
\ee
Analogously, for an aperture centered on a stellar peak we find
\be
\label{eq:tdeplstar}
\frac{\left[t_{\rm depl}\right]_{\rm star}}{\left[t_{\rm depl}\right]_{\rm gal}}=
\frac{\beta_{\rm g}\frac{t_{\rm over}}{t_{\rm gas}}\left[1+(\beta_{\rm g}-1)\frac{t_{\rm over}}{t_{\rm gas}}\right]^{-1}+\frac{t_{\rm star}}{\tau}\left(\frac{l_{\rm ap}}{\lambda}\right)^2}
       {1+\frac{t_{\rm star}}{\tau}\left(\frac{l_{\rm ap}}{\lambda}\right)^2} .
\ee
{In these equations, $\beta_{\rm g}\equiv\F_{\rm g,over}/\F_{\rm g,iso}$ indicates the ratio between the mean gas flux of peaks in the overlap $\F_{\rm g,over}$ and the mean flux of those in isolation $\F_{\rm g,over}$. Likewise, $\beta_{\rm s}\equiv\F_{\rm s,over}/\F_{\rm s,iso}$ indicates the same ratio for stellar fluxes. These flux ratios can be directly measured from observations if the spatial resolution allows the smallest apertures to contain only a single region (i.e.~$l_{\rm ap}<\lambda$). By only considering the smallest apertures, one can then obtain $\beta_{\rm g}$ and $\beta_{\rm s}$ by dividing the mean flux in apertures containing both tracers by the mean flux in those containing only a single tracer. If the spatial resolution is insufficient, some parametrization of the flux evolution needs to be assumed. For instance, if the gas (stellar) flux decreases to zero (increases from zero) linearly during the overlap and is constant otherwise, then $\beta_{\rm g}=\beta_{\rm s}=0.5$. The advantage of measuring the flux ratio rather than adopting some parametrization of the flux evolution is that equations~(\ref{eq:tdeplgas}) and~(\ref{eq:tdeplstar}) become independent of any prior assumptions.

In equations~(\ref{eq:tdeplgas}) and~(\ref{eq:tdeplstar}), the number 1 in the numerator or denominator indicates the guaranteed gas or stellar peak, respectively. The terms containing $(l_{\rm ap}/\lambda)^2$ represent the gas and stellar peaks residing in the aperture by chance. If $l_{\rm ap}\gg\lambda$, then both equations approach unity and the bias of the gas depletion time-scale vanishes. Finally, the term in the denominator (numerator) containing $\beta_{\rm s}$ ($\beta_{\rm g}$) reflects the non-zero probability of finding stars (gas) in the central gas (stellar) peak in case the gas and stellar phases overlap (i.e.~$t_{\rm over}\neq 0$).} As explained in \S\ref{sec:specific}, $t_{\rm over}$ encompasses the duration of SF $t_{\rm SF}$ as well as the time-scale for the removal of gas from the aperture or region by feedback $t_{\rm fb}=\min{(l_{\rm ap},\lambda)}/v_{\rm ej}$, where $v_{\rm ej}$ is the characteristic removal velocity of the gas, {be it by a phase transition or by motion}. Together, this yields $t_{\rm over}=t_{\rm SF}+\min{(l_{\rm ap},l_{\rm T})}/v_{\rm ej}$. When using a tracer that only detects unembedded stars, then $t_{\rm SF}$ must be close to zero and hence the duration of the overlap is mainly set by gas removal.

\begin{figure}
\center\resizebox{\hsize}{!}{\includegraphics{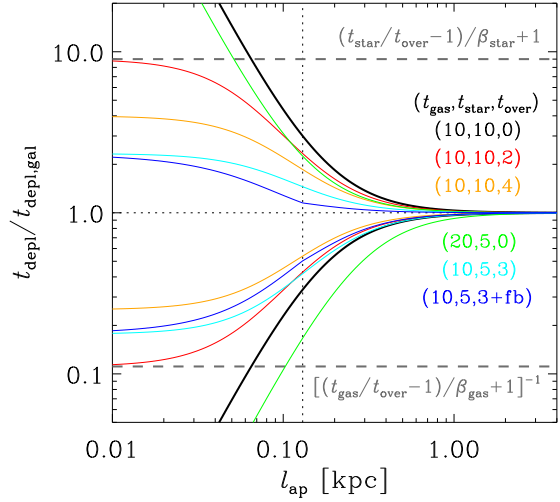}}\\
\caption[]{\label{fig:tdepl}
      {Expected relative change of the measured gas depletion time-scale as a function of aperture size when centering on gas peaks (top curves) or stellar peaks (bottom curves) for several combinations of $\{t_{\rm gas},t_{\rm star},t_{\rm over}\}$ (see legend). {Except where noted otherwise, gas removal due to feedback occurs instantaneously and hence the overlap equals the duration of SF ($t_{\rm over}=t_{\rm SF}$}). The thick black curves represent the reference variable set, with $t_{\rm gas}=t_{\rm star}$ and $t_{\rm over}=0$. The orange and red curves illustrate the effect of a non-zero overlap time, the green curve shows the effect of the ratio between the durations of the gas and stellar phases, and the cyan curve indicates the combined effect. {The blue curve adds the effect of non-instantaneous gas removal due to feedback, with a SF time-scale of $t_{\rm SF}=3$~Myr and a characteristic gas removal velocity of $v_{\rm ej}=100~{\rm km}~{\rm s}^{-1}$.} The vertical dotted line denotes the adopted typical separation between regions $\lambda=0.13$~kpc (cf.~Table~\ref{tab:examples}). {The (asymptotic) values reached for $l_{\rm ap}\downarrow0$ are indicated towards the right-hand side of the figure in grey. As can be verified by substituting equation~(\ref{eq:dxsamp}) into equations~(\ref{eq:tdeplgas}) and~(\ref{eq:tdeplstar}), the bias of $t_{\rm depl}$ never exceeds a factor of two as long as $l_{\rm ap}\geq\Delta x_{\rm samp}$.} }
                 }
\end{figure}
In Figure~\ref{fig:tdepl}, we show the relative change of the gas depletion time-scale resulting from equations~(\ref{eq:tdeplgas}) and~(\ref{eq:tdeplstar}) for different combinations of $t_{\rm gas}$, $t_{\rm star}$ and $t_{\rm over}$. While Figure~\ref{fig:scatter} already showed substantial scatter on scales smaller than a few 100~pc, we now see that centering an aperture on an overdensity of gas or stars systematically biases the gas depletion time-scale by up to an order of magnitude or more. {This is caused because focussing on a certain tracer guarantees it to be present in the aperture, which leads to a different depletion time-scale than measured on average throughout the galaxy. If $t_{\rm gas}=t_{\rm star}$, this occurs symmetrically around the galactic value of the depletion time, but in all other cases there exists an asymmetry between the curves focussing on gas or stars, which depends on the ratio $t_{\rm gas}/t_{\rm star}$. This is easily understood, because the bias is caused by the guarantee of having a gaseous or stellar peak -- if the visibility time-scale of either tracer exceeds that of the other, it will also be more numerous at a given instant. Therefore, the guarantee of including at least one region bright in that tracer will change the depletion time-scale by an amount smaller than when focussing on the shorter-lived tracer.

If the gas and SF tracers never overlap ($t_{\rm over}=0$), the depletion time-scale in the limit $l_{\rm ap}\downarrow0$ goes to infinity (zero) when focussing on gaseous (stellar) peaks. By contrast, a non-zero overlap between the gas and stellar phases (i.e.~the duration of SF {and gas removal}) introduces a flattening of the curves in Figure~\ref{fig:tdepl} for aperture sizes smaller than the typical separation of independent regions. When both phases overlap for some duration, then there is always a non-zero probability that both tracers are present in the aperture, even when only a single region is covered. This prevents the depletion time-scale from approaching zero or infinity for small aperture sizes.} In particular, when focussing on gas peaks we see that the bias of $t_{\rm depl}$ saturates at a value of
\be
\label{eq:asympgas}
\lim_{l_{\rm ap}\downarrow0}\frac{\left[t_{\rm depl}\right]_{\rm gas}}{\left[t_{\rm depl}\right]_{\rm gal}}=1+\beta_{\rm star}^{-1}\left(\frac{t_{\rm star}}{t_{\rm over}}-1\right) ,
\ee
whereas when focussing on stellar peaks it approaches
\be
\label{eq:asympstar}
\lim_{l_{\rm ap}\downarrow0}\frac{\left[t_{\rm depl}\right]_{\rm star}}{\left[t_{\rm depl}\right]_{\rm gal}}=\left[1+\beta_{\rm gas}^{-1}\left(\frac{t_{\rm gas}}{t_{\rm over}}-1\right)\right]^{-1} .
\ee
Because $t_{\rm star}$ is known from stellar population modelling, this implies that the relevant time-scales of the SF process can simply be read off figures like Figure~\ref{fig:tdepl}. {This is a potentially very powerful application of our framework, which is discussed in more detail below and in Appendix~\ref{sec:appcode}. Figure~\ref{fig:tdepl} also shows that the inclusion of a finite gas ejection velocity causes the flattening for small apertures to become more gradual, which occurs because the non-instantaneous removal of the gas increases the duration of the overlap for large aperture sizes (see above).

Note that the time-scales obtained through this method should be interpreted carefully. The time sequence of Figure~\ref{fig:tschem} follows the mass flow of the gas towards and through SF, i.e.~it is `Lagrangian'. Because not all of the gas is consumed in the SF process, a mass unit likely completes the sequence of Figure~\ref{fig:tschem} multiple times, either by only peripherally participating in the SF process or by actually forming a star and subsequently being ejected by feedback. In this context, $t_{\rm gas}$ reflects the {\it total} time {\it in between} subsequent SF events during which the mass unit is visible in the gas tracer. This also means that $t_{\rm gas}$ may span interruptions due to phase transitions unrelated to SF. While the duration of these interruptions themselves does not contribute to $t_{\rm gas}$, any prior visibility of a mass unit in the adopted gas tracer is contained in $t_{\rm gas}$.}

\begin{figure}
\center\resizebox{\hsize}{!}{\includegraphics{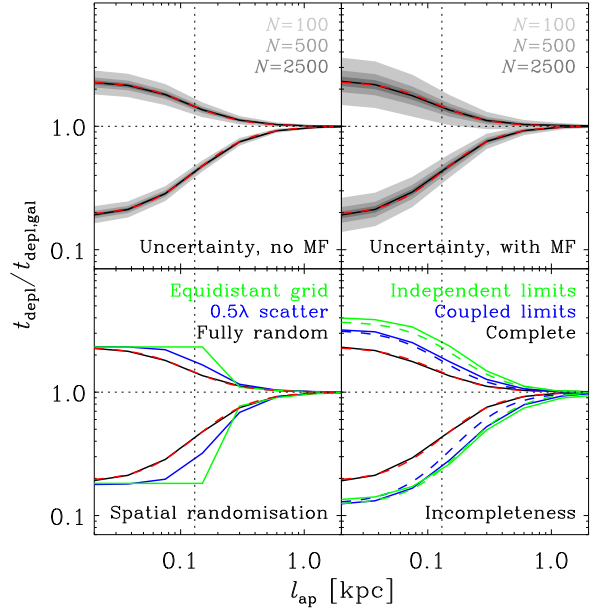}}\\
\caption[]{\label{fig:biasmc}
      {Comparison of the gas depletion time-scales seen in a simple Monte-Carlo experiment (see text) to those of the analytic expressions of equations~(\ref{eq:tdeplgas}) and~(\ref{eq:tdeplstar}), centering the apertures on gas peaks (top curves) or stellar peaks (bottom curves). {\it Top left}: The red dashed line indicates the analytic expression. The grey areas indicate the uncertainty range for small numbers of apertures (see legend) as obtained from the Monte-Carlo experiment. {\it Top right}: As in the previous panel, but assigning a power-law mass spectrum to the regions, covering two decades in mass with a slope of $-1.7$. {\it Bottom left}: Effect of the spatial randomisation, comparing the analytic expression (red dashed line) to the Monte-Carlo experiment for a hexagonal equidistant grid (green), for an additional random scatter up to $0.5\lambda$ (blue), and for a fully random distribution (black). {\it Bottom right}: Effect of incompleteness, comparing the analytic expression (red dashed line) to the Monte-Carlo experiment (black), using the same mass spectrum as in the top right panel. In the case of `coupled limits', the gas and stellar mass spectra are undetected below twice their minimum mass, whereas in the case of `independent limits' the gas (stellar) mass spectrum is undetected below five (three) times the minimum mass. Dashed lines indicate the analytic model when setting $t_{\rm gas}$, $t_{\rm star}$ and $t_{\rm over}$ to the mean time-scales for which the tracers are {\it detectable} (or detectably overlapping).}
                 }
\end{figure}
{As a first test of equations~(\ref{eq:tdeplgas})--(\ref{eq:asympstar}), we have performed a set of Monte-Carlo experiments very similar to those discussed in \S\ref{sec:scatter}. We randomly distribute 20,000 points over an area such that their mean separation is $\lambda=130$~pc and position each region randomly on the time sequence of Figure~\ref{fig:tschem}, using time-scales $\{t_{\rm gas},t_{\rm star},t_{\rm over}\}=\{10,5,3\}$~Myr. The gas (stellar) luminosity is taken to decrease (increase) linearly during the overlap phase while remaining constant when only a single tracer is present, implying $\beta_{\rm g}=\beta_{\rm s}=0.5$. We then place apertures of different sizes, which are focussed on each of these regions. The entire gas and stellar flux for the subsets of apertures centered on gas and stellar peaks, respectively, are then added up to obtain the bias of the gas depletion time-scale as a function of aperture size. In addition to this standard model, in some cases we assign a mass spectrum to the regions, account for detection limits, or consider different spatial distributions of the points.

Figure~\ref{fig:biasmc} shows the comparison between the Monte-Carlo model and the analytic expressions of equations~(\ref{eq:tdeplgas}) and~(\ref{eq:tdeplstar}). The values of $\beta_{\rm g}$ and $\beta_{\rm s}$ used in the analytic expressions are measured from the Monte-Carlo model as they would be determined from observed galaxies (see the earlier discussion) -- the other variables are simply set according to the initial conditions of the Monte-Carlo model. The top-left panel shows that when using only 100 apertures, an uncertainty of $\sim0.1$~dex should be expected when using the model to interpret observational data -- for larger numbers of apertures, theory and simulation converge to high accuracy. The scatter increases by about a factor of two when including a mass spectrum (top-right panel), to $\sim0.2$~dex, suggesting that the analysis proposed here requires a minimum of about 100 apertures to yield statistically useful results. Note that the applicability of the analytic expressions is unaffected by the presence of a mass spectrum. The third panel in Figure~\ref{fig:biasmc} addresses the assumption made thus far that independent regions are randomly distributed. If they are distributed on an equidistant, hexagonal grid with inter-point separation $\lambda$, the familiar saturation of the depletion time-scale at small aperture sizes is already attained at $\lambda$ -- at smaller size-scales there is never more than a single region residing in the aperture. The addition of random perturbations to this fixed distribution of points shifts the saturation point to $0.5\lambda$. In either case, the saturation value is still a reliable measure of $t_{\rm gas}$ and $t_{\rm over}$. 

The bottom-right panel of Figure~\ref{fig:biasmc} considers the effect of incompleteness -- the partial detection of the mass spectra of gaseous and stellar regions. Unsurprisingly, the bias of the gas depletion time-scale changes when only part of the mass spectrum is detected. This gives incorrect results when using the model to derive $t_{\rm gas}$ and $t_{\rm over}$ from observations. It makes no difference whether the detection limits of the gas and SF tracers are coupled (i.e.~the gas limit turns into the stellar limit and the number of regions is conserved between both tracers) or independent. However, we also show the analytic model when adopting the mean {\it detectability} time-scales of the gas and SF tracers rather than their underlying lifetimes, as well as the mean duration of their detectable overlap (dashed lines). These agree well with the Monte-Carlo experiment, indicating that the time-scales that are obtained from incomplete observations refer to the time-scales during which the tracers are {\it detectable}. When the observations are incomplete, this naturally differs from the underlying, true lifetimes of the gas and SF tracers and their overlap.}

We conclude that measurements of the gas depletion time-scale in small apertures centered on gas or SF tracers directly probe the duration of the several phases of the SF process. This may open up a new avenue to infer the physics of SF and feedback as a function of the galactic environment.

\section{Discussion} \label{sec:disc}
\subsection{Summary}
{We have presented a simple uncertainty principle for spatially resolved galactic SF relations. This explains the failure of these relations on small spatial scales as the result of the incomplete statistical sampling of independent star-forming regions. The main conclusions of this work are as follows.
\begin{enumerate}
\item Throughout most of the known star-forming systems, the incomplete sampling of independent star-forming regions determines the spatial scale $\Delta x$ below which galactic SF relations break down. It provides a more stringent criterion than the incomplete sampling of SF tracers from the IMF (which dominates in the outer regions of galaxy discs) or spatial drift (which dominates in galaxy centres). If the Toomre length sets the separation of independent star-forming regions, we predict that there should be little variation of $\Delta x$ as a function of the galactic gas surface density.
\item The Poisson statistics of sampling independent star-forming regions cause the scatter of the gas depletion time-scale of the spatially resolved SF relation to depend on the aperture size as $\sigma_{\log{t}}\propto l_{\rm ap}^{-\gamma}$ with $\gamma=0.5$--$1$ for $l_{\rm ap}=0.1$--1~kpc. The increase of the scatter with decreasing size-scale flattens for small aperture sizes, where it is dominated by the cloud mass spectrum and the details of the luminosity evolution during the SF process. We find good agreement with the observed dependence of the scatter on the spatial scale.
\item When focussing apertures on gas or stellar peaks, the measured gas depletion time-scale is biased to larger or smaller values, respectively. This bias directly probes the time-scales governing the SF process, such as the duration of the gas phase and its time overlap with the stellar phase. These time-scales can be obtained from galaxy-wide observations without the need to spatially resolve independent star-forming regions -- resolving their mean separation suffices. Simple Monte-Carlo models of large numbers of star-forming regions show that the method is insensitive to the cloud mass spectrum and can be applied reliably when at least 100 gas or stellar peaks are used. Another important strength of this method is that it is not hampered by the uncertain conversion factors between gas tracer flux and gas mass that traditionally plague the inference of SF physics from galactic SF relations \citep[see e.g.][]{kennicutt12}, because it relies exclusively on tracer flux ratios.
\end{enumerate}
Fortran and IDL modules for applying the uncertainty principle are available at http://www.mpa-garching.mpg.de/KL14principle. A checklist detailing the required steps for their observational application is supplied in Appendix~\ref{sec:appcode}.}

\subsection{Assumptions, observational caveats and biases}
The statistical arguments used to derive the expected scatter in SF relations rely on several implicit assumptions. If this theoretical framework is applied in regimes where these assumptions break down, the derived scatter will vary from that predicted. For example, we have assumed that the galactic SFR is roughly constant over $\tau$, which will break down in a localised `starburst' event.\footnote{{In the framework of this paper, the clouds and regions in a galaxy-wide starburst are no longer independent, which implies that the entire galaxy actually constitutes a single independent region.}} Alternatively, if the physical properties of a galaxy vary {substantially} within a given observational aperture, the characteristic size and mass of independent regions may also change, potentially giving rise to additional scatter. Also, the above framework has been defined under the simplest assumption that the galaxy is face-on. As inclination will directly affect several key variables (e.g.~the projected aperture area, gas/star surface density, rotation curves), the deprojected values should be used.

Our framework also assumes that observations accurately recover the full underlying gas and young star distributions. In the idealised case where all independent star-forming regions have the same characteristic mass (e.g.~the Toomre mass $M_{\rm T}=\sigma^4/G^2\Sigma$), the observations need to be sufficiently sensitive to both (1) detect this characteristic gas mass and (2) detect the stellar population {eventually} resulting from this gas (e.g.~the Toomre mass times some SF efficiency). However, in practice the gas mass distribution will be continuous and hence sensitivity limitations mean that observations will only detect emission from a fraction of the gas and young stars. {We have shown in \S\ref{sec:physics} that this affects the time-scales that are retrieved when considering the bias of the gas depletion time-scale in small apertures. Ideally, the gas and SF tracer observations are complete, but at least they should encompass the same total fraction of gas clouds and resulting young stellar populations they produce.} A robust application of our framework requires the recovered fraction to be large. The requirement that observations be sensitive to a similar, substantial fraction of the gas and young stars being produced effectively places a distance limit on the applicability of this analysis. {These and other observational points of caution are discussed in more detail in Appendix~\ref{sec:checklist}.}

One final thing to bear in mind, is that our uncertainty principle mainly considers the statistics of observational tracers of a physical system rather than directly describing that system itself. {As described in \S\ref{sec:physics}, the physics of SF can be characterised by considering their effects on these statistics.}

\subsection{Implications and future applications} \label{sec:future}
The uncertainty principle presented in this paper shows that care must be taken when comparing galactic SF relations to small-scale SF relations and provides quantitative limits on their applicability. For instance, SF relations measured in the solar neighbourhood \citep{heiderman10,lada10,gutermuth11,lada14} are fundamentally different from their galactic counterparts. Small-scale SF relations describe the conversion of dense (and likely self-gravitating) gas into stars, whereas large-scale SF relations additionally cover galactic physics such as feedback, cooling and inflow dynamics. Despite this added complexity, the clear advantage is a better statistical sampling of the SF process -- and as shown in this paper, the small-scale characteristics of the SF process can be obtained by considering how the large-scale SF relations break down.

The framework of this paper also shows that {there is no reason why the SF recipes that are used in high-resolution numerical simulations of galaxy formation and evolution \citep[e.g.][]{springel03} should be motivated by the galactic SF relation, which in essence fails to describe the $\Delta x\ll 100$~pc size-scales that can actually be resolved in modern calculations}. {Because the presented framework is Lagrangian in nature (i.e.~it traces the mass flow through a system), it may also be used as a means to quickly analyse mass flows in Eulerian, grid-based simulations, alleviating the need for tracer particles.}

The dependence of the scatter and bias of the gas depletion time-scale on the aperture size (see Figures~\ref{fig:scatter} and~\ref{fig:tdepl}) presents a novel and potentially powerful way of deriving the durations of the different phases of the SF process, such as how long the gas and stars are visible in their respective tracers, and how long the phase lasts during which both are visible (i.e.~the duration of SF {itself as well as the resulting gas removal}). Any possible degeneracies can be lifted by considering different tracers. For instance, when using H$\alpha$ to trace stars and assuming $t_{\rm star}=2$~Myr, $\Delta x_{\rm samp}\propto(\tau/\Delta t)^{1/2}$ is indistinguishable for $t_{\rm gas}=\{1,4\}$~Myr. However, when adding FUV and hence $t_{\rm star}=14$~Myr, both gas tracer lifetimes can be separated and the degeneracy is lifted. {The use of different SF tracers that capture the early, embedded phase of SF and are sensitive to different stellar masses (such as cm continuum, {far-infrared} and young stellar object counts) may be used to map the assembly of the stellar mass function.} Similarly, the combination of different gas tracers in our framework can be used to constrain time-scales for gas phase transitions (e.g.~H{\sc i}$\rightarrow$H$_2$) and to map the gas volume density evolution towards SF as a function of absolute time. We conclude that our uncertainty principle and its corresponding framework provide a powerful tool to characterise the SF process, using spatially resolved, galaxy-scale observations.

{Finally, we note that the statistical model presented in this paper is very general and applies to any astrophysical process that can be separated into (partially overlapping) subsequent phases: any system that is subject to some degree of time-evolution. By contrast, it cannot be applied when the correlated quantities are the simultaneous result of an underlying phenomenon (e.g.~the colour-magnitude relation of main sequence stars). The fundamental point is that when a macroscopic correlation is caused by a time-evolution, then it {\it must} break down on small scales because the subsequent phases are resolved. This general observation supports the application of the presented framework to a wide range of fields, from small-scale star and planet formation to galaxy formation and evolution.}

\section*{Acknowledgements}
{We thank an anonymous referee for a helpful report. We are very grateful to Andreas Schruba for helpful discussions, advice and detailed comments on the manuscript, and to Frank Bigiel, Simon Glover, Adam Leroy, Andreas Burkert and Ralf Klessen for stimulating discussions and/or comments on an early draft of this work.} We acknowledge the Aspen Center for Physics for their hospitality and to the National Science Foundation for support, Grant No.~1066293.

\bibliographystyle{mn2e}
\bibliography{mybib}

\begin{thebibliography}{41}
\expandafter\ifx\csname natexlab\endcsname\relax\def\natexlab#1{#1}\fi
\small
\bibitem[{{Bigiel} {et~al}\mbox{.}(2010){Bigiel}, {Leroy}, {Walter}, {Blitz},
  {Brinks}, {de Blok}, \& {Madore}}]{bigiel10}
{Bigiel} F., {Leroy} A., {Walter} F., {Blitz} L., {Brinks} E., {de Blok}
  W.~J.~G., {Madore} B., 2010, \aj, 140, 1194

\bibitem[{{Bigiel} {et~al}\mbox{.}(2008){Bigiel}, {Leroy}, {Walter}, {Brinks},
  {de Blok}, {Madore}, \& {Thornley}}]{bigiel08}
{Bigiel} F., {Leroy} A., {Walter} F., {Brinks} E., {de Blok} W.~J.~G., {Madore}
  B., {Thornley} M.~D., 2008, \aj, 136, 2846

\bibitem[{{Bigiel} {et~al}\mbox{.}(2011){Bigiel}, {Leroy}, {Walter}, {Brinks},
  {de Blok}, {Kramer}, {Rix}, {Schruba}, {Schuster}, {Usero}, \&
  {Wiesemeyer}}]{bigiel11}
{Bigiel} F. {et~al.}, 2011, \apjl, 730, L13

\bibitem[{{Blanc} {et~al}\mbox{.}(2009){Blanc}, {Heiderman}, {Gebhardt},
  {Evans}, \& {Adams}}]{blanc09}
{Blanc} G.~A., {Heiderman} A., {Gebhardt} K., {Evans}, II N.~J., {Adams} J.,
  2009, \apj, 704, 842

\bibitem[{{Burkert} \& {Hartmann}(2012)}]{burkert13}
{Burkert} A., {Hartmann} L., 2012, \apj~submitted, arXiv:1212.4543

\bibitem[{{Calzetti}, {Liu} \& {Koda}(2012){Calzetti}, {Liu}, \&
  {Koda}}]{calzetti12}
{Calzetti} D., {Liu} G., {Koda} J., 2012, \apj, 752, 98

\bibitem[{{Chabrier}(2003)}]{chabrier03}
{Chabrier} G., 2003, \pasp, 115, 763

\bibitem[{{Elmegreen}(1997)}]{elmegreen97b}
{Elmegreen} B.~G., 1997, in Revista Mexicana de Astronomia y Astrofisica, vol.
  27, Vol.~6, Revista Mexicana de Astronomia y Astrofisica Conference Series,
  {Franco} J., {Terlevich} R., {Serrano} A., eds., p. 165

\bibitem[{{Elmegreen}(2002)}]{elmegreen02}
{Elmegreen} B.~G., 2002, \apj, 577, 206

\bibitem[{{Feldmann}, {Gnedin} \& {Kravtsov}(2011){Feldmann}, {Gnedin}, \&
  {Kravtsov}}]{feldmann11}
{Feldmann} R., {Gnedin} N.~Y., {Kravtsov} A.~V., 2011, \apj, 732, 115

\bibitem[{{Fumagalli}, {da Silva} \& {Krumholz}(2011){Fumagalli}, {da Silva},
  \& {Krumholz}}]{fumagalli11}
{Fumagalli} M., {da Silva} R.~L., {Krumholz} M.~R., 2011, \apjl, 741, L26

\bibitem[{{Genzel} {et~al}\mbox{.}(2010){Genzel}, {Tacconi}, {Gracia-Carpio},
  {Sternberg}, {Cooper}, {Shapiro}, {Bolatto}, {Bouch{\'e}}, {Bournaud},
  {Burkert}, {Combes}, {Comerford}, {Cox}, {Davis}, {Schreiber},
  {Garcia-Burillo}, {Lutz}, {Naab}, {Neri}, {Omont}, {Shapley}, \&
  {Weiner}}]{genzel10}
{Genzel} R. {et~al.}, 2010, \mnras, 407, 2091

\bibitem[{{Gutermuth} {et~al}\mbox{.}(2011){Gutermuth}, {Pipher}, {Megeath},
  {Myers}, {Allen}, \& {Allen}}]{gutermuth11}
{Gutermuth} R.~A., {Pipher} J.~L., {Megeath} S.~T., {Myers} P.~C., {Allen}
  L.~E., {Allen} T.~S., 2011, \apj, 739, 84

\bibitem[{{Heiderman} {et~al}\mbox{.}(2010){Heiderman}, {Evans}, {Allen},
  {Huard}, \& {Heyer}}]{heiderman10}
{Heiderman} A., {Evans}, II N.~J., {Allen} L.~E., {Huard} T., {Heyer} M., 2010,
  \apj, 723, 1019

\bibitem[{{Heisenberg}(1927)}]{heisenberg27}
{Heisenberg} W., 1927, Zeitschrift fur Physik, 43, 172

\bibitem[{{Kennicutt} \& {Evans}(2012)}]{kennicutt12}
{Kennicutt} R.~C., {Evans} N.~J., 2012, \araa, 50, 531

\bibitem[{{Kennicutt}(1998)}]{kennicutt98b}
{Kennicutt}, Jr. R.~C., 1998, \apj, 498, 541

\bibitem[{{Kruijssen}(2012)}]{kruijssen12d}
{Kruijssen} J.~M.~D., 2012, \mnras, 426, 3008

\bibitem[{{Kruijssen} {et~al}\mbox{.}(2013){Kruijssen}, {Longmore},
  {Elmegreen}, {Murray}, {Bally}, {Testi}, \& {Kennicutt, Jr.}}]{kruijssen13}
{Kruijssen} J.~M.~D., {Longmore} S.~N., {Elmegreen} B.~G., {Murray} N., {Bally}
  J., {Testi} L., {Kennicutt, Jr.} R.~C., 2013, \mnras~submitted,
  arXiv:1303.6286

\bibitem[{{Kruijssen} {et~al}\mbox{.}(2014){Kruijssen}, {Schruba}, {Longmore},
  \& {Bigiel}}]{kruijssen14}
{Kruijssen} J.~M.~D., {Schruba} A., {Longmore} S.~N., {Bigiel} F., 2014, in
  prep. (K14)

\bibitem[{{Krumholz} \& {McKee}(2005)}]{krumholz05}
{Krumholz} M.~R., {McKee} C.~F., 2005, \apj, 630, 250

\bibitem[{{Lada} {et~al}\mbox{.}(2013){Lada}, {Lombardi}, {Roman-Zuniga},
  {Forbrich}, \& {Alves}}]{lada14}
{Lada} C., {Lombardi} M., {Roman-Zuniga} C., {Forbrich} J., {Alves} J., 2013,
  \apj~in~press, arXiv:1309.7055

\bibitem[{{Lada}, {Lombardi} \& {Alves}(2010){Lada}, {Lombardi}, \&
  {Alves}}]{lada10}
{Lada} C.~J., {Lombardi} M., {Alves} J.~F., 2010, \apj, 724, 687

\bibitem[{{Lee} {et~al}\mbox{.}(2011){Lee}, {Gil de Paz}, {Kennicutt},
  {Bothwell}, {Dalcanton}, {Jos{\'e} G.~Funes S.}, {Johnson}, {Sakai},
  {Skillman}, {Tremonti}, \& {van Zee}}]{lee11}
{Lee} J.~C. {et~al.}, 2011, \apjs, 192, 6

\bibitem[{{Leitherer} {et~al}\mbox{.}(1999){Leitherer}, {Schaerer}, {Goldader},
  {Delgado}, {Robert}, {Kune}, {de Mello}, {Devost}, \&
  {Heckman}}]{leitherer99}
{Leitherer} C. {et~al.}, 1999, \apjs, 123, 3

\bibitem[{{Leroy} {et~al}\mbox{.}(2012){Leroy}, {Bigiel}, {de Blok},
  {Boissier}, {Bolatto}, {Brinks}, {Madore}, {Munoz-Mateos}, {Murphy},
  {Sandstrom}, {Schruba}, \& {Walter}}]{leroy12}
{Leroy} A.~K. {et~al.}, 2012, \aj, 144, 3

\bibitem[{{Leroy} {et~al}\mbox{.}(2008){Leroy}, {Walter}, {Brinks}, {Bigiel},
  {de Blok}, {Madore}, \& {Thornley}}]{leroy08}
{Leroy} A.~K., {Walter} F., {Brinks} E., {Bigiel} F., {de Blok} W.~J.~G.,
  {Madore} B., {Thornley} M.~D., 2008, \aj, 136, 2782

\bibitem[{{Leroy} {et~al}\mbox{.}(2013){Leroy}, {Walter}, {Sandstrom},
  {Schruba}, {Munoz-Mateos}, {Bigiel}, {Bolatto}, {Brinks}, {de Blok}, {Meidt},
  {Rix}, {Rosolowsky}, {Schinnerer}, {Schuster}, \& {Usero}}]{leroy13}
{Leroy} A.~K. {et~al.}, 2013, \apj~in~press, arXiv:1301.2328

\bibitem[{{Liu} {et~al}\mbox{.}(2011){Liu}, {Koda}, {Calzetti}, {Fukuhara}, \&
  {Momose}}]{liu11}
{Liu} G., {Koda} J., {Calzetti} D., {Fukuhara} M., {Momose} R., 2011, \apj,
  735, 63

\bibitem[{{Longmore} {et~al}\mbox{.}(2013){Longmore}, {Bally}, {Testi},
  {Purcell}, {Walsh}, {Bressert}, {Pestalozzi}, {Molinari}, {Ott}, {Cortese},
  {Battersby}, {Murray}, {Lee}, {Kruijssen}, {Schisano}, \&
  {Elia}}]{longmore13}
{Longmore} S.~N. {et~al.}, 2013, \mnras, 429, 987

\bibitem[{{Longmore} {et~al}\mbox{.}(2014){Longmore}, {Kruijssen}, {Bastian},
  {Bally}, {Rathborne}, {Testi}, {Stolte}, {Dale}, {Bressert}, \&
  {Alves}}]{longmore14}
{Longmore} S.~N. {et~al.}, 2014, Protostars and Planets VI, in press

\bibitem[{{Martin} \& {Kennicutt}(2001)}]{martin01}
{Martin} C.~L., {Kennicutt}, Jr. R.~C., 2001, \apj, 555, 301

\bibitem[{{Onodera} {et~al}\mbox{.}(2010){Onodera}, {Kuno}, {Tosaki}, {Kohno},
  {Nakanishi}, {Sawada}, {Muraoka}, {Komugi}, {Miura}, {Kaneko}, {Hirota}, \&
  {Kawabe}}]{onodera10}
{Onodera} S. {et~al.}, 2010, \apjl, 722, L127

\bibitem[{{Schmidt}(1959)}]{schmidt59}
{Schmidt} M., 1959, \apj, 129, 243

\bibitem[{{Schruba} {et~al}\mbox{.}(2011){Schruba}, {Leroy}, {Walter},
  {Bigiel}, {Brinks}, {de Blok}, {Dumas}, {Kramer}, {Rosolowsky}, {Sandstrom},
  {Schuster}, {Usero}, {Weiss}, \& {Wiesemeyer}}]{schruba11}
{Schruba} A. {et~al.}, 2011, \aj, 142, 37

\bibitem[{{Schruba} {et~al}\mbox{.}(2010){Schruba}, {Leroy}, {Walter},
  {Sandstrom}, \& {Rosolowsky}}]{schruba10}
{Schruba} A., {Leroy} A.~K., {Walter} F., {Sandstrom} K., {Rosolowsky} E.,
  2010, \apj, 722, 1699

\bibitem[{{Silk}(1997)}]{silk97}
{Silk} J., 1997, \apj, 481, 703

\bibitem[{{Springel} \& {Hernquist}(2003)}]{springel03}
{Springel} V., {Hernquist} L., 2003, \mnras, 339, 289

\bibitem[{{Thilker} {et~al}\mbox{.}(2007){Thilker}, {Bianchi}, {Meurer}, {Gil
  de Paz}, {Boissier}, {Madore}, {Boselli}, {Ferguson}, {Mu{\~n}oz-Mateos},
  {Madsen}, {Hameed}, {Overzier}, {Forster}, {Friedman}, {Martin}, {Morrissey},
  {Neff}, {Schiminovich}, {Seibert}, {Small}, {Wyder}, {Donas}, {Heckman},
  {Lee}, {Milliard}, {Rich}, {Szalay}, {Welsh}, \& {Yi}}]{thilker07}
{Thilker} D.~A. {et~al.}, 2007, \apjs, 173, 538

\bibitem[{{Wolfire} {et~al}\mbox{.}(2003){Wolfire}, {McKee}, {Hollenbach}, \&
  {Tielens}}]{wolfire03}
{Wolfire} M.~G., {McKee} C.~F., {Hollenbach} D., {Tielens} A.~G.~G.~M., 2003,
  \apj, 587, 278

\bibitem[{{Yusef-Zadeh} {et~al}\mbox{.}(2009){Yusef-Zadeh}, {Hewitt}, {Arendt},
  {Whitney}, {Rieke}, {Wardle}, {Hinz}, {Stolovy}, {Lang}, {Burton}, \&
  {Ramirez}}]{yusefzadeh09}
{Yusef-Zadeh} F. {et~al.}, 2009, \apj, 702, 178

\end{thebibliography}

\appendix
\onecolumn
\section{IDL and Fortran routines for applying the uncertainty principle} \label{sec:appcode}
\subsection{Code description}
Fortran and IDL modules for applying the uncertainty principle are available at http://www.mpa-garching.mpg.de/KL14principle. The modules contain functions to predict the following quantities.
\begin{enumerate}
\item The size-scale $\Delta x$ below which galactic SF relations fail.
\item The logarithmic scatter of the depletion time-scale (see Appendix~\ref{sec:appscatter}).
\item The relative change of the depletion time-scale when focussing the aperture on gas or stellar peaks (see Appendix~\ref{sec:appbias}).
\end{enumerate}
The functions are written in a general form which does not necessarily refer to the SF process. They all require some choice of the duration of the first phase $t_1$ (e.g.~$t_{\rm gas}$), the second phase $t_2$ (e.g.~$t_{\rm star}$), their overlap $t_{\rm over}$ and the separation $\lambda$, and are accompanied by comprehensive documentation and examples.

\subsection{Checklist for observational applications} \label{sec:checklist}
We focus here on the specific steps needed to determine $t_{\rm gas}$, $t_{\rm star}$ and $t_{\rm over}$ from observed galaxies using the framework of this paper. In particular, certain input parameters need to be estimated and it should be verified that the assumptions made for equations~(\ref{eq:tdeplgas}) and~(\ref{eq:tdeplstar}) hold.
\begin{enumerate}
\item The gas and SF tracer maps should satisfy a number of simple criteria. (a) Contamination should be minimized, i.e.~the tracers should be directly associated with the physical objects that they are intended to probe. Possible sources of contamination include degenerate tracers (e.g.~24$\mu$m emission traces both young and evolved stars) and the scatter of photons at large distances from their original sources. (b) The maps should be as complete as possible, i.e.~the loss of flux due to extinction, leakage, excitation, and chemistry should be minor, and if the gas tracer map relies on interferometric data, it should always be combined with single-dish observations. (c) The sensitivity of the gas and SF tracer observations should be sufficient to recover a large fraction of the gas and SF tracer emission. (d) The spatial resolution of the maps should be sufficient to resolve the separation between independent star-forming regions. Individual regions do not need to be resolved -- in fact, point (ii) shows that the analysis is simplified when they are not. (e) One of the three time-scales $t_{\rm gas}$, $t_{\rm star}$ and $t_{\rm over}$ should be known a priori. For most practical applications, the known time-scale will be $t_{\rm star}$, because a stellar population synthesis model \citep[e.g.][]{leitherer99} can be used to estimate over which time-scale $t_{\rm star}$ the SF tracer probes the young stellar emission.
\item A sample of gas and SF tracer peaks can be identified using a clump-finding algorithm. In this paper, we assumed the star-forming regions to be much smaller than their separation (i.e.~they are treated as point particles), whereas in reality the ISM forms a continuous structure. This introduces a certain subjectivity in the selection process. While there is no intrinsically preferred method, ideally the identification should ensure that independent star-forming regions are identified as single clumps rather than being divided into smaller fragments of which the positions on the time sequence of Figure~\ref{fig:tschem} may be correlated. However, with no prior definition of the size-scale on which the peaks become independent, it is unclear how the identified peaks actually relate to independent regions. This uncertainty affects the method of \S\ref{sec:physics} in two ways. Firstly, the separation between independent regions $\lambda$ (see point [iii]) may be estimated incorrectly. Secondly, if a single independent region is broken up into several peaks, this will bias the measured time-scales. For instance, if a gaseous independent region without SF tracer emission is broken up into multiple peaks, separate apertures will be centered on each of these peaks and the gas depletion time will be biased to larger values. When $t_{\rm star}$ is known, this causes $t_{\rm over}$ and $t_{\rm gas}$ to be underestimated. The measured time-scales may therefore depend on the selection criteria (beyond their plausible physical variation with e.g.~the cloud mass) and hence the analysis should be performed for a range of selection criteria to test their robustness. A more detailed example of how to tackle this problem will be presented in K14. A comparison with the method of \S\ref{sec:physics} may also be used to test the selection criteria (e.g.~by verifying that the bias of the gas depletion time-scale never exceeds a factor of two for $l_{\rm ap}\geq\Delta x_{\rm samp}$).
\item The typical separation between star-forming regions is measured by counting the gas and SF peaks (making sure that overlapping gaseous and stellar peaks are only counted once) and calculating $\lambda=2\sqrt{A/\pi N}$, where $A$ is the total area of the observed field. Note that this is a geometric mean and does not account for the possible clustering of regions. Also, if the spatial distribution of regions is fully random, the expected distance to each point's nearest neighbour is smaller than $\lambda$.
\item The overlap-to-isolated flux ratios $\beta_{\rm g}\equiv\F_{\rm g,over}/\F_{\rm g,iso}$ and $\beta_{\rm s}\equiv\F_{\rm s,over}/\F_{\rm s,iso}$ can be directly measured from observations if the spatial resolution allows the smallest apertures to contain only a single region (i.e.~$l_{\rm ap}<\lambda$). By only considering the smallest apertures, one can then obtain $\beta_{\rm g}$ and $\beta_{\rm s}$ by dividing the mean flux in apertures containing both tracers by the mean flux in those apertures containing only a single tracer. If the spatial resolution is insufficient, some parametrization of the flux evolution needs to be assumed (see \S\ref{sec:physics}).
\item The bias of the gas depletion time-scale at an aperture size $l_{\rm ap}$ is then calculated by centering at least 100 artificial apertures\footnote{The method works with a smaller number of apertures, but the uncertainties will increase due to low-number statistics (see \S\ref{sec:physics} and Figure~\ref{fig:biasmc}).} of size $l_{\rm ap}$ on (1) gas or (2) SF tracer peaks. For these two separate cases, the entire gas and SF tracer fluxes should be added up across all apertures. The biases when focussing on gas or SF tracer peaks are obtained by dividing the resulting gas-to-SF tracer flux ratios by the galactic mean.
\item The size-scale dependence of the bias of the gas depletion time-scale is obtained by repeating the above for different aperture sizes. The time-scale {\it ratios} $t_{\rm star}/t_{\rm over}$ and $t_{\rm gas}/t_{\rm over}$ are then measured by doing a $\chi^2$ minimization of equations~(\ref{eq:tdeplgas}) and~(\ref{eq:tdeplstar}) while varying both ratios. If one of the three time-scales is known (in most cases $t_{\rm star}$ is obtained from stellar population modelling, see point [i]), the time-scale ratios are easily converted to absolute values (in most cases $t_{\rm gas}$ and $t_{\rm over}$). If the gas or SF tracer maps are incomplete, these time-scales reflect the {\it detectability} time-scales, which may differ from the underlying physical time-scales. Therefore, it is desirable to use high-sensitivity observations that recover a large fraction of the gas and SF tracer flux.
\end{enumerate}
An important strength of this method is that it is not hampered by the uncertain conversion factors between gas tracer flux and gas mass that traditionally plague the inference of SF physics from galactic SF relations \citep[see e.g.][]{kennicutt12}, because it relies exclusively on tracer flux ratios.

\section{Derivation of the scatter of SF relations} \label{sec:appscatter}
This appendix details the derivation of the four scatter terms in equation~(\ref{eq:scatter}), and in particular the scatter due to the Poisson sampling of independent star-forming regions.

The gas depletion time-scale measured in a given aperture is defined as $t_{\rm depl}\equiv M_{\rm gas}/{\rm SFR}\propto \F_{\rm gas}/\F_{\rm SF}$, where $\F_{\rm gas}$ and $\F_{\rm SF}$ indicate the flux emitted within the aperture by gas and SF tracers, respectively. Because we are interested in the scatter in $\log{t_{\rm depl}}$, the proportionality constant is irrelevant. The fluxes can be specified by counting the number of expected regions for each particular phase and multiplying them by the corresponding mean flux received from a single region. For this, we need to identify the numbers of isolated gaseous (i.e.~no SF tracer emission is present) regions $N_{\rm g,iso}$, isolated stellar regions $N_{\rm s,iso}$, and overlap (i.e.~both gas and SF tracer emission are present) regions $N_{\rm over}$. Because we are considering randomly-placed apertures, the total number of regions in the aperture is $N_{\rm rnd}=(l_{\rm ap}/\lambda)^2$, where the subscript `rnd' refers to `random' and indicates that these regions are randomly positioned on the time sequence of Figure~\ref{fig:tschem}. This leads to the definitions:
\bea
\label{eq:appn}
N_{\rm g,iso}\equiv\frac{t_{\rm gas}-t_{\rm over}}{\tau}\left(\frac{l_{\rm ap}}{\lambda}\right)^2 ;
N_{\rm s,iso}\equiv\frac{t_{\rm star}-t_{\rm over}}{\tau}\left(\frac{l_{\rm ap}}{\lambda}\right)^2 ;
N_{\rm over}\equiv\frac{t_{\rm over}}{\tau}\left(\frac{l_{\rm ap}}{\lambda}\right)^2 .
\eea
These are expectation values in the absence of further contraints -- the true number depends on which combinations of $N_{\rm g,iso}$, $N_{\rm s,iso}$ and $N_{\rm over}$ are allowed. When calculating the scatter in $\log{t_{\rm depl}}$, only those apertures are included that contain emission from both the gas and SF tracer. In other words, the combinations
\bea
\label{eq:appcombos}
(1)~N_{\rm g,iso}=0, N_{\rm s,iso}=0, N_{\rm over}=0 ;
(2)~N_{\rm g,iso}>0, N_{\rm s,iso}=0, N_{\rm over}=0 ;
(3)~N_{\rm g,iso}=0, N_{\rm s,iso}>0, N_{\rm over}=0 ,
\eea
are excluded. The number distributions considered here follow Poisson statistics, in which the probability of drawing a number $k$ at an expectation value $N$ is given by $p(k,N)=N^k\exp{(-N)}/k!$. The probabilities of the above three combinations are thus
\bea
\label{eq:appprobs}
p_1=\e^{-N_{\rm g,iso}-N_{\rm s,iso}-N_{\rm over}} ;
p_2=\e^{-N_{\rm s,iso}-N_{\rm over}}\left(1-\e^{-N_{\rm g,iso}}\right) ;
p_3=\e^{-N_{\rm g,iso}-N_{\rm over}}\left(1-\e^{-N_{\rm s,iso}}\right) ,
\eea
where the two terms in parentheses indicate the probabilities of having non-zero numbers of isolated gaseous and stellar regions, respectively.

We can now express the proportionality of the gas depletion time-scale in terms of $N_{\rm g,iso}$, $N_{\rm s,iso}$ and $N_{\rm over}$, omitting the three disallowed combinations:
\bea
\label{eq:apptdepl}
t_{\rm depl}\propto\frac{\F_{\rm gas}}{\F_{\rm star}}&=&\frac
{N_{\rm over}\F_{\rm g,over}+N_{\rm g,iso}\F_{\rm g,iso}-p_2N_{\rm g,iso}\left(1-\e^{-N_{\rm g,iso}}\right)^{-1}\F_{\rm g,iso}}
{N_{\rm over}\F_{\rm s,over}+N_{\rm s,iso}\F_{\rm s,iso}-p_3N_{\rm s,iso}\left(1-\e^{-N_{\rm s,iso}}\right)^{-1}\F_{\rm s,iso}} \nonumber\\
&=&\left[\frac
{\beta_{\rm g}N_{\rm over}+N_{\rm g,iso}\left(1-\e^{-N_{\rm s,iso}-N_{\rm over}}\right)}
{\beta_{\rm s}N_{\rm over}+N_{\rm s,iso}\left(1-\e^{-N_{\rm g,iso}-N_{\rm over}}\right)}\right]
\frac{\F_{\rm g,iso}}{\F_{\rm s,iso}} .
\eea
In the first equality, the terms including $p_2$ and $p_3$ subtract the disallowed combinations from the otherwise expected gas and SF tracer fluxes. They include the terms in parentheses to correctly reflect the expected numbers of non-zero gaseous (combination~2) or stellar (combination~3) regions, respectively. By disallowing the three combinations of equation~(\ref{eq:appcombos}), the total probability no longer adds up to unity, but to $1-p_1-p_2-p_3$. The numerator and the denominator represent the expected gaseous and SF tracer flux, respectively, and hence should each be divided by the total probability. This correction is identical for the numerator and the denominator and therefore it cancels.

Going back to \S\ref{sec:scatter}, we defined the scatter in $\log{t_{\rm depl}}$ as
\be
\label{eq:appscatter}
\sigma_{\log{t}}^2=\sigma_{\rm samp}^2+\sigma_{\rm evo}^2+\sigma_{\rm MF}^2+\sigma_{\rm obs}^2 ,
\ee
where $\sigma_{\rm samp}$ indicates the Poisson error, $\sigma_{\rm evo}$ represents the scatter caused by the luminosity evolution of independent regions, $\sigma_{\rm MF}$ is the scatter due to the mass spectrum, and $\sigma_{\rm obs}$ denotes the intrinsic observational error. In the context of equation~(\ref{eq:apptdepl}), $\sigma_{\rm samp}$ is set by the variances of $N_{\rm g,iso}$, $N_{\rm s,iso}$ and $N_{\rm over}$. Formally, $\sigma_{\rm evo}$ follows similarly from the variances of $\beta_{\rm g}$, $\beta_{\rm s}$, $\F_{\rm g,iso}(t)$ and $\F_{\rm s,iso}(t)$ (i.e.~their time-evolution), and $\sigma_{\rm MF}$ is set by the variances of $\F_{\rm g,iso}(0)$ and $\F_{\rm s,iso}(0)$ (i.e.~their instantaneous variances, which are caused by the underlying mass spectrum), but both are kept as free parameters here because they depend strongly on the properties of the system under consideration (see below and \S\ref{sec:scatter}). The observational error $\sigma_{\rm obs}$ is also a free parameter, which has to be determined separately for each observation.

Following the above separation of the scatter into four different components, we define $\sigma_{\rm samp}$ as
\be
\label{eq:appsamp}
\sigma_{\rm samp}^2=\alpha^2\left[\left(\frac{\dau\ln{t_{\rm depl}}}{\dau N_{\rm g,iso}}\right)^2\sigma_{N_{\rm g,iso}}^2+
\left(\frac{\dau\ln{t_{\rm depl}}}{\dau N_{\rm s,iso}}\right)^2\sigma_{N_{\rm s,iso}}^2+
\left(\frac{\dau\ln{t_{\rm depl}}}{\dau N_{\rm over}}\right)^2\sigma_{N_{\rm over}}^2\right] ,
\ee
where $\alpha\equiv1/\ln{10}\approx0.43$ converts the logarithmic scatter from base $\e$ to base $10$. The covariance terms are omitted on purpose, because the added complexity is not warranted by the precision gained (see Figure~\ref{fig:scatter} and \S\ref{sec:scatter}). The three variances are Poissonian and thus read
\bea
\label{eq:apppoisson}
\sigma_{N_{\rm g,iso}}=N_{\rm g,iso}^{1/2} ;
\sigma_{N_{\rm s,iso}}=N_{\rm s,iso}^{1/2} ;
\sigma_{N_{\rm over}}=N_{\rm over}^{1/2} .
\eea
The three derivatives are obtained by differentiation of equation~(\ref{eq:apptdepl}):
\bea
\label{eq:appder}
\frac{\dau\ln{t_{\rm depl}}}{\dau N_{\rm g,iso}}&=&
\frac{1-\e^{-N_{\rm s,iso}-N_{\rm over}}}{\beta_{\rm g}N_{\rm over}+N_{\rm g,iso}\left(1-\e^{-N_{\rm s,iso}-N_{\rm over}}\right)}-
\frac{N_{\rm s,iso}\e^{-N_{\rm g,iso}-N_{\rm over}}}{\beta_{\rm s}N_{\rm over}+N_{\rm s,iso}\left(1-\e^{-N_{\rm g,iso}-N_{\rm over}}\right)} ;\nonumber\\
\frac{\dau\ln{t_{\rm depl}}}{\dau N_{\rm s,iso}}&=&
\frac{N_{\rm g,iso}\e^{-N_{\rm s,iso}-N_{\rm over}}}{\beta_{\rm g}N_{\rm over}+N_{\rm g,iso}\left(1-\e^{-N_{\rm s,iso}-N_{\rm over}}\right)}-
\frac{1-\e^{-N_{\rm g,iso}-N_{\rm over}}}{\beta_{\rm s}N_{\rm over}+N_{\rm s,iso}\left(1-\e^{-N_{\rm g,iso}-N_{\rm over}}\right)} ;\\
\frac{\dau\ln{t_{\rm depl}}}{\dau N_{\rm over}}&=&
\frac{\beta_{\rm g}+N_{\rm g,iso}\e^{-N_{\rm s,iso}-N_{\rm over}}}{\beta_{\rm g}N_{\rm over}+N_{\rm g,iso}\left(1-\e^{-N_{\rm s,iso}-N_{\rm over}}\right)}-
\frac{\beta_{\rm s}+N_{\rm s,iso}\e^{-N_{\rm g,iso}-N_{\rm over}}}{\beta_{\rm s}N_{\rm over}+N_{\rm s,iso}\left(1-\e^{-N_{\rm g,iso}-N_{\rm over}}\right)} .\nonumber
\eea
Together, equations~(\ref{eq:appn}),~(\ref{eq:appsamp}),~(\ref{eq:apppoisson}) and~(\ref{eq:appder}) define $\sigma_{\rm samp}$ in equation~(\ref{eq:appscatter}).

The terms in equation~(\ref{eq:appscatter}) accounting for the scatter due to the gas and SF tracer luminosity evolution and the mass spectrum of independent regions represent errors of the mean. They therefore depend both on the actual numbers of regions contained in the aperture and on the scatter induced for a single region ($\sigma_{\rm evo,1g}$, $\sigma_{\rm evo,1s}$ and $\sigma_{{\rm MF},1}$, indicating the scatter for a single region due to the gaseous luminosity evolution, the stellar luminosity evolution, and the mass spectrum, respectively):
\bea
\label{eq:appevomf}
\sigma_{\rm evo}^2=\sigma_{{\rm evo},1g}^2N_{\rm gas}^{-1}+\sigma_{{\rm evo},1s}^2N_{\rm star}^{-1} ; \sigma_{\rm MF}=\sigma_{{\rm MF},1}N_{\rm tot}^{-1/2} ,
\eea
where $N_{\rm tot}$ is the total number of star-forming regions in the aperture, $N_{\rm gas}$ the total number of gaseous regions, and $N_{\rm star}$ the total number of stellar regions. Subtracting the disallowed combinations of equation~(\ref{eq:appcombos}), we obtain
\bea
\label{eq:appntot}
N_{\rm tot}&=&
\frac{N_{\rm g,iso}+N_{\rm s,iso}+N_{\rm over}-p_2N_{\rm g,iso}\left(1-\e^{-N_{\rm g,iso}}\right)^{-1}-p_3N_{\rm s,iso}\left(1-\e^{-N_{\rm s,iso}}\right)^{-1}}{1-p_1-p_2-p_3}\nonumber\\&=&
\frac{N_{\rm g,iso}\left(1-\e^{-N_{\rm s,iso}-N_{\rm over}}\right)+N_{\rm s,iso}\left(1-\e^{-N_{\rm g,iso}-N_{\rm over}}\right)+N_{\rm over}}{1-\e^{-N_{\rm g,iso}-N_{\rm s,iso}-N_{\rm over}}-\e^{-N_{\rm s,iso}-N_{\rm over}}\left(1-\e^{-N_{\rm g,iso}}\right)-\e^{-N_{\rm g,iso}-N_{\rm over}}\left(1-\e^{-N_{\rm s,iso}}\right)} ,
\eea
for the total number of regions and likewise
\bea
\label{eq:appngasnstar}
N_{\rm gas}&=&
\frac{N_{\rm g,iso}\left(1-\e^{-N_{\rm s,iso}-N_{\rm over}}\right)+N_{\rm over}}{1-\e^{-N_{\rm g,iso}-N_{\rm s,iso}-N_{\rm over}}-\e^{-N_{\rm s,iso}-N_{\rm over}}\left(1-\e^{-N_{\rm g,iso}}\right)-\e^{-N_{\rm g,iso}-N_{\rm over}}\left(1-\e^{-N_{\rm s,iso}}\right)}\\
N_{\rm star}&=&
\frac{N_{\rm s,iso}\left(1-\e^{-N_{\rm g,iso}-N_{\rm over}}\right)+N_{\rm over}}{1-\e^{-N_{\rm g,iso}-N_{\rm s,iso}-N_{\rm over}}-\e^{-N_{\rm s,iso}-N_{\rm over}}\left(1-\e^{-N_{\rm g,iso}}\right)-\e^{-N_{\rm g,iso}-N_{\rm over}}\left(1-\e^{-N_{\rm s,iso}}\right)} ,
\eea
for the actual numbers of gaseous and stellar regions contained in the aperture. For $l_{\rm ap}\downarrow0$, we have $\sigma_{\rm evo}^2\rightarrow\sigma_{{\rm evo},1g}^2+\sigma_{{\rm evo},1s}^2$, $\sigma_{\rm MF}\rightarrow\sigma_{{\rm MF},1}$ and $\sigma_{\rm samp}\downarrow0$, where $\sigma_{{\rm evo},1g}$, $\sigma_{{\rm evo},1s}$ and $\sigma_{{\rm MF},1}$ are free parameters (see \S\ref{sec:scatter}).

\section{Derivation of the relative change of the gas depletion time-scale} \label{sec:appbias}
This appendix details the derivation of equations~(\ref{eq:tdeplgas}) and~(\ref{eq:tdeplstar}). The gas depletion time-scale measured in a given aperture is defined as $t_{\rm depl}\equiv M_{\rm gas}/{\rm SFR}\propto \F_{\rm gas}/\F_{\rm SF}$, where $\F_{\rm gas}$ and $\F_{\rm SF}$ indicate the flux emitted within the aperture by gas and SF tracers, respectively. When focussing apertures of different sizes on gaseous or stellar peaks, the relative change (or bias) of the gas depletion time-scale with respect to the galactic average $[t_{\rm depl}]_{\rm gal}$ thus becomes
\be
\label{eq:appratio}
\frac{\left[t_{\rm depl}\right]}{\left[t_{\rm depl}\right]_{\rm gal}}=\frac{\F_{\rm gas}}{\F_{\rm star}}\frac{\F_{\rm s,gal}}{\F_{\rm g,gal}} ,
\ee
where $\F_{\rm g,gal}$ and $\F_{\rm s,gal}$ indicate the galaxy-integrated flux of the gas and SF tracers, respectively. The fluxes in this equation can be specified by counting the number of expected regions and multiplying them by the expected flux received from a single region. The expected flux from a gaseous region is
\be
\label{eq:app1gas}
\F_{\rm gas,1}=\frac{t_{\rm gas}-t_{\rm over}}{t_{\rm gas}}\F_{\rm g,iso}+\frac{t_{\rm over}}{t_{\rm gas}}\F_{\rm g,over}=\left[1+\frac{(\beta_{\rm g}-1)t_{\rm over}}{t_{\rm gas}}\right]\F_{\rm g,iso} ,
\ee
where $\F_{\rm g,iso}$ is the mean gas tracer flux of peaks in isolation (i.e.~no SF tracer emission is present), $\F_{\rm g,over}$ is that of peaks in the overlap, and $\beta_{\rm g}\equiv\F_{\rm g,over}/\F_{\rm g,iso}$. In the first equality, the ratios $(t_{\rm gas}-t_{\rm over})/t_{\rm gas}$ and $t_{\rm over}/t_{\rm gas}$ indicate the probabilities that a region is an isolated gas peak or resides in the overlap, respectively. Analogously, for a stellar region we have
\be
\label{eq:app1star}
\F_{\rm star,1}=\frac{t_{\rm star}-t_{\rm over}}{t_{\rm star}}\F_{\rm s,iso}+\frac{t_{\rm over}}{t_{\rm star}}\F_{\rm s,over}=\left[1+\frac{(\beta_{\rm s}-1)t_{\rm over}}{t_{\rm star}}\right]\F_{\rm s,iso} .
\ee
If we now define a galaxy to contain a total number of $N_{\rm tot}$ independent regions, a fraction $t_{\rm gas}/\tau$ of these will show gas tracer emission and a fraction $t_{\rm star}/\tau$ will show SF tracer emission. Hence, the galactic flux ratio between gas and SF tracers becomes
\be
\label{eq:appratiogal}
\frac{\F_{\rm g,gal}}{\F_{\rm s,gal}}=\frac{(t_{\rm gas}/\tau)N_{\rm tot}\F_{\rm gas,1}}{(t_{\rm star}/\tau)N_{\rm tot}\F_{\rm star,1}}=\frac{t_{\rm gas}\F_{\rm gas,1}}{t_{\rm star}\F_{\rm star,1}} ,
\ee
The above equations hold irrespective of whether the apertures are centered on gas of stellar peaks.

We now consider the gas and SF tracer fluxes emanating from an aperture of size $l_{\rm ap}$. Given a characteristic separation between regions $\lambda$, the number of random (either gas, stars, or both) regions in the aperture is $N_{\rm rnd}=(l_{\rm ap}/\lambda)^2$. When an aperture is focussed on a gas peak, the total number of gaseous regions thus becomes $N_{\rm gas}=1+(t_{\rm gas}/\tau)N_{\rm rnd}$ and we expect a gas flux from the aperture of
\be
\label{eq:appgasapgas}
\F_{\rm gas}=N_{\rm gas}\F_{\rm gas,1}=\left[1+\frac{(\beta_{\rm g}-1)t_{\rm over}}{t_{\rm gas}}\right]\left[1+\frac{t_{\rm gas}}{\tau}\left(\frac{l_{\rm ap}}{\lambda}\right)^2\right]\F_{\rm g,iso} .
\ee
In the same aperture, there is a probability $t_{\rm over}/t_{\rm gas}$ that the central gas peak also contains stars by residing in the overlap. In addition, a fraction $t_{\rm star}/\tau$ of the random regions in the aperture will show SF tracer emission. The total SF tracer flux from the aperture is therefore
\be
\label{eq:appgasapstar}
\F_{\rm star}=\frac{t_{\rm over}}{t_{\rm gas}}\F_{\rm s,over}+\frac{t_{\rm star}}{\tau}\left(\frac{l_{\rm ap}}{\lambda}\right)^2\F_{\rm star,1}=\left\{\beta_{\rm s}\frac{t_{\rm over}}{t_{\rm gas}}+\left[1+\frac{(\beta_{\rm s}-1)t_{\rm over}}{t_{\rm star}}\right]\frac{t_{\rm star}}{\tau}\left(\frac{l_{\rm ap}}{\lambda}\right)^2\right\}\F_{\rm s,iso} .
\ee
Substituting equations~(\ref{eq:appratiogal}),~(\ref{eq:appgasapgas}) and~(\ref{eq:appgasapstar}) into equation~(\ref{eq:appratio}) then yields equation~(\ref{eq:tdeplgas}), i.e.~the bias of the gas depletion time-scale in an aperture focussed on a gas tracer peak:
\be
\label{eq:apptdeplgas}
\frac{\left[t_{\rm depl}\right]_{\rm gas}}{\left[t_{\rm depl}\right]_{\rm gal}}=
\frac{1+(t_{\rm gas}/{\tau})({l_{\rm ap}}/{\lambda})^2}
       {\beta_{\rm s}({t_{\rm over}}/{t_{\rm star}})\left[1+(\beta_{\rm s}-1)({t_{\rm over}}/{t_{\rm star}})\right]^{-1}+(t_{\rm gas}/{\tau})({l_{\rm ap}}/{\lambda})^2} .
\ee
As $l_{\rm ap}\downarrow0$, the bias becomes entirely set by the term accounting for the possibility that the central gas tracer peak resides in the overlap. This is the reason that the bias of $t_{\rm depl}$ in focussed apertures can be used to derive the time-scales involved in the SF process.

Analogously to the above, an aperture focussed on a stellar peak contains a gas flux of
\be
\label{eq:appstarapgas}
\F_{\rm gas}=\frac{t_{\rm over}}{t_{\rm star}}\F_{\rm g,over}+\frac{t_{\rm gas}}{\tau}\left(\frac{l_{\rm ap}}{\lambda}\right)^2\F_{\rm gas,1}=\left\{\beta_{\rm g}\frac{t_{\rm over}}{t_{\rm star}}+\left[1+\frac{(\beta_{\rm g}-1)t_{\rm over}}{t_{\rm gas}}\right]\frac{t_{\rm gas}}{\tau}\left(\frac{l_{\rm ap}}{\lambda}\right)^2\right\}\F_{\rm g,iso} ,
\ee
and a total SF tracer flux of
\be
\label{eq:appstarapstar}
\F_{\rm star}=N_{\rm star}\F_{\rm star,1}=\left[1+\frac{(\beta_{\rm s}-1)t_{\rm over}}{t_{\rm star}}\right]\left[1+\frac{t_{\rm star}}{\tau}\left(\frac{l_{\rm ap}}{\lambda}\right)^2\right]\F_{\rm s,iso} .
\ee
Substituting equations~(\ref{eq:appratiogal}),~(\ref{eq:appstarapgas}) and~(\ref{eq:appstarapstar}) into equation~(\ref{eq:appratio}) we obtain equation~(\ref{eq:tdeplstar}), i.e.~the bias of the gas depletion time-scale in an aperture focussed on a SF tracer peak:
\be
\label{eq:apptdeplstar}
\frac{\left[t_{\rm depl}\right]_{\rm star}}{\left[t_{\rm depl}\right]_{\rm gal}}=
\frac{\beta_{\rm g}({t_{\rm over}}/{t_{\rm gas}})\left[1+(\beta_{\rm g}-1)({t_{\rm over}}/{t_{\rm gas}})\right]^{-1}+(t_{\rm star}/{\tau})({l_{\rm ap}}/{\lambda})^2}
{1+(t_{\rm star}/{\tau})({l_{\rm ap}}/{\lambda})^2} ,
\ee
which does not depend on any of the involved fluxes and for $l_{\rm ap}\downarrow0$ becomes entirely set by the term accounting for the possibility that the central SF tracer peak resides in the overlap.

\bsp

\label{lastpage}

\end{document}